\documentclass[final,authoryear,3p,times,twocolumn]{elsarticle}
\usepackage[latin2]{inputenc}
\usepackage{graphicx}
\usepackage{amssymb}
\usepackage{amsthm}
\usepackage{ulem}

\journal{Applied Radiation and Isotopes 114(2016)128}

\begin{document}

\begin{frontmatter}



\title{Activation cross sections of proton induced nuclear reactions on palladium up to 80 MeV}


\author[1]{F. T\'ark\'anyi}
\author[1]{F. Ditr\'oi\corref{*}}
\author[1]{S. Tak\'acs}
\author[1] {J. Csikai}
\author[2]{A. Hermanne}
\author[3]{S. Uddin}
\author[3]{M. Baba}
%\author[4]{A.V. Ignatyuk}

\cortext[*]{Corresponding author: ditroi@atomki.hu}

\address[1]{Institute for Nuclear Research, Hungarian Academy of Sciences (ATOMKI),  Debrecen, Hungary}
\address[2]{Cyclotron Laboratory, Vrije Universiteit Brussel (VUB), Brussels, Belgium}
\address[4]{Institute of Physics and Power Engineering (IPPE), Obninsk, Russia}
\address[3]{Cyclotron Radioisotope Center (CYRIC), Tohoku University, Sendai, Japan}

\begin{abstract}
\noindent Activation cross sections of proton induced nuclear reactions on palladium were measured up to 80 MeV by using the stacked foil irradiation technique and gamma ray spectrometry. The beam intensity, the incident energy and the energy degradation were controlled by a method based on flux constancy via normalization to the excitation functions of monitor reactions measured in parallel. Excitation functions for direct and cumulative cross-sections were measured for the production of ${}^{104m,104g,105}$${}^{g,106m,110m}$Ag, ${}^{100,101}$Pd, ${}^{99m,99g,100,}$${}^{101m}$${}^{,101g,102m,102g,105}$Rh and ${}^{103,}$${}^{97}$Ru radioisotopes. The cross section data were compared with the theoretical predictions of TENDL-2014 and -2015 libraries. For practical applications thick target yields were derived from the measured excitation functions. Application in the field of medical radionuclide production is shortly discussed.
\end{abstract}


\end{frontmatter}

% \linenumbers

\section{Introduction}
\label{1}
Excitation functions of proton induced nuclear reactions on palladium are important in many applications including medical radioisotope production, determination of elemental impurities by activation analysis, development of low activation materials and others (Khandaker et al., 2010; Skakun and Rauscher, 2007). At present only very few experimental data exist in the literature for light ion induced nuclear reactions on this material. The present investigations were initiated with the aim to investigate the production possibilities of some medically related radioisotopes (${}^{104}$Ag, ${}^{101}$Rh, ${}^{103}$Pd) and to produce reliable activation curves for thin layer activation (${}^{105}$Ag). We already reported the activation cross sections of proton, deuteron, ${}^{3}$He and alpha-particle induced reactions on Pd for these three radionuclides \citep{DF2012, DF2007, Hermanne2004, Hermanne2005, IAEATLA, TF2009}.  Regarding the proton activation data, our earlier experiments were performed in 2002-2004 and at that moment only a few earlier low energy work on (p,n) reactions were found in the literature \citep{Batii, Hu1998, Johnson, Kormali, Zarubin}.  Our results obtained in that period for some medically relevant radionuclides were published during 2004-2007. Since then a detailed study of Pd activation by protons up to 40 MeV was published by \citep{Khandaker2010}, low energy data for formation of ${}^{105}$Ag were published by \citep{Dillmann2006, Dillmann2011}. As we had the opportunity to extend the experimental cross sections for all reaction products formed in proton irradiation on Pd up to 80 MeV, we decided to report our new results complemented with all our unpublished data.

\section{Experimental}
\label{2}

\noindent The irradiations were carried out at the external beam lines of the AVF cyclotron at the Tohoku University, Japan at 80 MeV and the CGR 560 cyclotron at the Vrije Universiteit Brussels at 37.3, 28.5 and 15.9 MeV irradiation energies. High purity Pd foils were stacked with Al, Cu and Ti monitor foils and irradiated for 0.1--1.0 hours at 200 nA nominal beam intensity. All foils were purchased from Goodfellow {\copyright}, England and were at least 99.7\% pure. The beam intensity, the incident energy and the energy degradation were controlled by a method based on flux constancy via normalization (including fitting) to the excitation functions of ${}^{27}$Al(p,\textit{x})${}^{22,24}$Na and ${}^{nat}$Cu(p,x)${}^{56,58}$Co,${}^{62,65}$Zn and ${}^{nat}$Ti(p,x)${}^{48}$V  monitor reactions re-measured in parallel over  the whole investigated energy range (see Table 1)(T\'{a}rk\'{a}nyi et al., 2001). Measurement of the gamma-ray spectra of the activated foils was carried out without chemical separation. The measurements were started from a few hours up to several weeks after EOB.  

\noindent The beam currents were deduced from the monitor reactions, the number of the target nuclei from the precise mass and surface determination of the target foils. The calculated (by using home-made code based on \citep{Andersen} energy degradation of the incident beam was checked and corrected on the basis of the monitor reactions. For calculation of the activity of the produced radioisotopes the decay data (Table 3) were taken from \citep{Nudat}.

\noindent The uncertainties of the cross-sections were estimated from the linearly contributing processes supposing equal sensitivities. The total resulting errors were between 8 \% and 13 \%. In the estimation of the uncertainty of the medium proton energy in each target foil in addition to the well-known beam energy broadening, the cumulative propagation of the uncertainty in the energy of the incident beam and in the foil thickness were taken into account. 

\noindent The experimental and data evaluation details for our different irradiations are collected in Table1 and Table 2, the used decay data and the Q-values of the contributing reactions in Table 3.

\begin{table*}[t]
\tiny
\caption{Main parameters of the experiments and the methods of data evaluations}
\begin{center}

\begin{tabular}{|p{0.5in}|p{0.6in}|p{0.6in}|p{0.6in}|p{0.6in}|p{0.6in}|p{0.6in}|p{0.6in}|} \hline 
\multicolumn{8}{|c|}{\textbf{\eject Experiment}} \\ \hline 
Reaction & \multicolumn{7}{|p{4.2in}|}{ ${}^{nat}$Pd(p,x)} \\ \hline 
Method  & \multicolumn{7}{|p{4.2in}|}{Stacked foil} \\ \hline 
Irradiation(year) & 2002 & 2002 & 2003 & 2004 & 2002 & 2002 & 2002 \\ \hline 
Identification & S2002-1 & S2002-2 & S2003 & S2004 & V1 & V2 & V3 \\ \hline 
Target stack and thicknesses \newline ($\mu$m) & Cu 54.4\newline Pd 12\newline Al 97.8\newline Pt 9.29\newline Al 97.8\newline Ta 8.63\newline block\newline \newline \newline \newline repeated\newline 9 times & Cu 54.4\newline Ta 209.7\newline Pd 9.5\newline Al 97.8\newline Fe 25.4\newline block\newline \newline \newline \newline \newline repeated\newline 10 times & Cu 54.4\newline Al 1000\newline Pd 12.1\newline Cd 15\newline Al 1000\newline block\newline \newline \newline \newline \newline repeated\newline 10 times & Cu 54.4\newline Mo 49.6\newline Y 110.4\newline Pd 12.5\newline Ag 303\newline Cd 15\newline Al 495 or 1000\newline block\newline \newline repeated\newline 10 times & Cu 23.92\newline Ti 12.02\newline Pd 7.99\newline \newline \newline \newline \newline \newline \newline \newline repeated\newline 9  times & Ni 23.97\newline Pd 7.99\newline Ti 30.58\newline Ta 8.63\newline \newline \newline \newline \newline \newline \newline repeated\newline 10 times & Ni 23.97\newline Pd 7.99\newline Cu 23.92\newline Al 100\newline Ti 30.58\newline Pd 7.99\newline Ta 8.63\newline Al 100\newline \newline \newline repeated\newline 10 times \\ \hline 
Number of  Pd target foils & 9 & 10 & 10 & 10 & 9 & 10 & 10 \\ \hline 
Accelerator & K110 MeV cyclotron\newline  CYRIC, Sendai & K110 MeV cyclotron\newline  CYRIC, Sendai & K110 MeV cyclotron\newline  CYRIC, Sendai & K110 MeV cyclotron\newline  CYRIC, Sendai & CGR-560 cyclotron\newline VUB\newline Brussels & CGR-560 cyclotron\newline VUB\newline Brussels & CGR-560 cyclotron\newline VUB\newline Brussels \\ \hline 
Primary energy (MeV) & 70($\pm$0.3) & 70($\pm$0.3) & 70($\pm$0.3) & 80($\pm$0.3) & 15.9($\pm$0.2) & 28.5($\pm$0.2) & 37.3($\pm$0.2) \\ \hline 
Energy range covered (for Pd)\newline (MeV) & 69.5-28.9 & 64.4-42.8 & 67.7-34.1 & 79.9-37.4 & 15.8-6.7 & 28.2-15.9 & 37.0-20.0 \\ \hline 
Irradiation time (min) & 60 & 60 & 6.4 & 71 & 30 & 30 & 30 \\ \hline 
Beam current (nA) & 90  & 100  & 160 & 49.6  & 170.4 & 183.7  & 175.5  \\ \hline 
Monitor reaction, [recommended values]  & $^{27}$Al(p,x)$^{22,24}$Na\newline $^{nat}$Cu(p,x)$^{62,65}$Zn\newline  & $^{27}$Al(p,x)$^{22,24}$Na\newline $^{nat}$Cu(p,x)$^{62,65}$Zn\newline  & $^{27}$Al(p,x)$^{22,24}$Na\newline $^{nat}$Cu(p,x)$^{62,65}$Zn\newline  & $^{27}$Al(p,x)$^{22,24}$Na\newline $^{nat}$Cu(p,x)$^{62,65}$Zn\newline  & $^{nat}$Cu(p,x)$^{62,65}$Zn\newline $^{nat}$Ti(p,x)$^{48}$V\newline  & $^{nat}$Cu(p,x)$^{62,65}$Zn\newline $^{nat}$Ti(p,x)$^{48}$V\newline  & $^{nat}$Cu (p,x)$^{62,65}$Zn\newline $^{nat}$Ti(p,x)$^{48}$V\newline  \\ \hline 
Detector & HPGe & HPGe & HPGe & HPGe & HPGe & HPGe & HPGe \\ \hline 
$\gamma$-spectra measurements & 3 series & 3 series & 2 series & 2 series & 3 series & 3 series & 3 series \\ \hline 
Cooling times (h) & 55.1-57.1\newline 87.0-102.9\newline 562-658 & 37.0-39.2\newline 105.3-109.4\newline 920-993 & 0.7-1.4\newline 34.1-67.7 & 20.7-23.8\newline 74.1-98.9 & 0.7-2.3\newline 27.0-30.4\newline 31.8-150.1 & 0.7-2.7\newline 40.1-49.8\newline 49.4-194.8 & 1.7-5.0\newline 24.8-31.9\newline 341.4-456.3 \\ \hline 
Earlier published\newline (from these experiments) (Ditr\'{o}i et al., 2007; Hermanne et al., 2004b; T\'{a}rk\'{a}nyi et al., 2009) & ${}^{105}$Ag\newline ${}^{106m}$Ag\newline ${}^{100}$Pd\newline ${}^{101m}$Rh & ${}^{105}$Ag\newline ${}^{106m}$Ag\newline ${}^{100}$Pd\newline ${}^{101m}$Rh & ${}^{103}$Ag,${}^{103}$Pd &  & ${}^{103}$Ag\newline ${}^{105}$Ag\newline ${}^{106m}$Ag\newline ${}^{104g}$Ag & ${}^{103}$Ag\newline ${}^{105}$Ag\newline ${}^{106m}$Ag\newline ${}^{104g}$Ag & ${}^{103}$Ag\newline ${}^{105}$Ag\newline ${}^{106m}$Ag\newline ${}^{110m}$Ag\newline ${}^{104g}$Ag \\ \hline 
\end{tabular}

\end{center}
\end{table*}

\begin{table*}[t]
\tiny
\caption{Methods of data evaluation}
\begin{center}
\begin{tabular}{|p{1.2in}|p{3.4in}|} \hline 
\multicolumn{2}{|p{1in}|}{\textbf{Data evaluation}} \\ \hline 
Method & Preliminary study of earlier experimental data and theoretical predictions \\ \hline 
Gamma spectra evaluation & Genie 2000, Forgamma \citep{Canberra, Szekely}, automatic and  manual controlled peak fitting \\ \hline 
Determination of beam intensity & Faraday cup (preliminary), Fitted monitor reaction (final) \citep{TF1991} (see Table 1) \\ \hline 
Decay data & NUDAT 2.6 \citep{Kinsey, Nudat}  Toi Lund  database \citep{Chu} for data missing in NUDAT \citep{BNL} \\ \hline 
Reaction Q-values & Q-value calculator \citep{Pritychenko} \\ \hline 
Determination of  beam energy & Andersen \citep{Andersen}(calculated, preliminary), Fitted monitor reactions (final) \\ \hline 
Uncertainty of energy & Cumulative effects of possible uncertainties (primary energy, foil thickness and uniformity, energy straggling) \\ \hline 
Cross sections & Elemental cross section \\ \hline 
Uncertainty of cross sections & Sum in quadrature of all individual contribution \citep{Error}: (uncertainty on beam current  \%, target thickness  \%, detector efficiency calibration \%, $\gamma$-ray abundance and statistical uncertainty on count rates \%) \newline The contributions of the uncertainties of non-linear parameters were neglected (cooling times, half-life). \\ \hline 
Yield & Calculated from cross sections Physical yield \citep{Bonardi} \\ \hline 
Theory & TALYS (TENDL-2014 and 2015) \citep{Koning2014} \\ \hline 
\end{tabular}

\end{center}
\end{table*}

\begin{table*}[t]
\tiny
\caption{Decay data of the investigated reaction products (\citep{Nudat, Pritychenko}}
\begin{center}
\begin{tabular}{|p{0.8in}|p{0.6in}|p{0.6in}|p{0.5in}|p{0.8in}|p{0.8in}|} \hline 
Nuclide\newline (isomeric state)\newline Decay path & Half-life & E${}_{?}$(keV) & I${}_{?}$(\%) & Contributing reaction & Q-value\newline (keV)\newline gs $\mathrm{\to}$ gs \\ \hline 
\textbf{${}^{110m}$Ag\newline }117.59\textit{5 }keV\textit{\newline }IT: 1.33~\%\newline ~$\betaup$${}^{-}$: 98.67 \%\textbf{} & 249.83 d & ~657.7600\newline 677.6217\newline ~706.6760\newline 763.9424\newline ~884.6781\newline 937.485\newline ~1384.2931\newline ~1505.0280 & 95.61\newline 10.70~\newline ~16.69\newline ~22.60\newline ~75.0\newline ~35.0\newline 25.1~\newline 13.33 & ${}^{110}$Pd(p,n) & -1656.11 \\ \hline 
\textbf{${}^{106m}$Ag\newline }89.66\textit{7 }keV\newline $\varepsilonup$: 100~\%\newline \textbf{} & ~8.28 d & 406.182\newline 429.646\newline 450.976~\newline 616.17~\newline 717.34\newline 748.36\newline ~804.28\newline ~~824.69\newline 1045.83\newline ~1128.02\newline ~1199.39\newline 1527.65 & 13.4~\newline ~13.2~\newline ~28.2\newline ~21.6\newline 28.9~\newline ~20.6~\newline ~12.4\newline ~15.3\newline ~29.6\newline ~11.8~\newline 11.2~\newline ~16.3 & ${}^{106}$Pd(p,n)\newline ${}^{108}$Pd(p,3n)\newline ${}^{110}$Pd(p,5n) & -3747.49\newline -19507.09\newline -34456.89\newline  \\ \hline 
\textbf{${}^{105g}$Ag\newline }$\varepsilonup$: 100~\%\textbf{} & ~41.29 d & 280.44\newline 331.51\newline 344.52\newline 442.25\newline 443.37\newline ~644.55 & 30.2 \newline ~4.10\newline 41.4\newline ~4.72~\newline 10.5\newline 11.1 & ${}^{105}$Pd(p,n)\newline ${}^{106}$Pd(p,2n)\newline ${}^{108}$Pd(p,4n)\newline ${}^{110}$Pd(p,6n) & -2129.28\newline -11690.24\newline -27449.84\newline -42399.65\newline \newline  \\ \hline 
\textbf{${}^{104m}$Ag\newline }6.90\textit{22}~keV\newline IT: 0.07 \%\newline ~EC: 36.93~\%\newline $\betaup$${}^{+}$: 63 \%~\textbf{} & 33.5 min & 996.1\newline 1238.8 & ~0.50~\newline ~3.9 & ${}^{104}$Pd(p,n)\newline ${}^{105}$Pd(p,2n)\newline ${}^{106}$Pd(p,3n)\newline ${}^{108}$Pd(p,5n)\newline ${}^{110}$Pd(p,7n) & -5061.0\newline -12155.1\newline -21716.08\newline -37475.67\newline -52425.48 \\ \hline 
\textbf{${}^{104g}$Ag\newline }~EC: 85~\%\newline $\betaup$${}^{+}$: 15 \%~\textbf{\newline } & 69.2 min & ~623.2~\newline 863.0\newline ~925.9~\newline ~941.6 & ~2.5\newline 6.9\newline 12.5\newline ~25.0 & ${}^{104}$Pd(p,n)\newline ${}^{105}$Pd(p,2n)\newline ${}^{106}$Pd(p,3n)\newline ${}^{108}$Pd(p,5n)\newline ${}^{110}$Pd(p,7n) & -5061.0\newline -12155.1\newline -21716.08\newline -37475.67\newline -52425.48 \\ \hline 
\textbf{${}^{103}$Ag\newline }EC: 73~\%~\newline $\betaup$${}^{+}$: 27 \%~\textbf{} & 65.7 min & 118.74\newline 148.20\newline 266.86\newline ~1273.83 & 31.2\newline ~28.3\newline 13.3\newline ~9.4 & ${}^{104}$Pd(p,2n)\newline ${}^{105}$Pd(p,3n)\newline ${}^{106}$Pd(p,4n)\newline ${}^{108}$Pd(p,6n)\newline ${}^{110}$Pd(p,8n) & -13448.35\newline -20542.46\newline -30103.43\newline -45863.02\newline -45863.02 \\ \hline 
\textbf{${}^{103}$Pd\newline }$\varepsilonup$: 100~\%\textbf{\newline } & 16.991 d & 357.45 & ~0.0221 & ${}^{104}$Pd(p,pn)\newline ${}^{105}$Pd(p,p2n)\newline ${}^{106}$Pd(p,p3n)\newline ${}^{108}$Pd(p,p5n)\newline ${}^{110}$Pd(p,p7n)\newline ${}^{103}$Ag decay & -9981.26\newline -17075.36\newline -26636.34\newline -42395.93\newline -57345.72\newline -13448.35 \\ \hline 
\textbf{${}^{101}$Pd\newline }EC: 92.8~\%\newline $\betaup$${}^{+}$: 7.2 \%~\textbf{} & 8.47 h & 269.67\newline 296.29\newline 565.98\newline 590.44 & 6.43\newline 19\newline 3.44\newline 12.06\newline  & ${}^{102}$Pd(p,pn)\newline ${}^{104}$Pd(p,p3n)\newline ${}^{105}$Pd(p,p4n)\newline ${}^{106}$Pd(p,p5n)\newline ${}^{108}$Pd(p,p7n)\newline ${}^{110}$Pd(p,p9n)\newline ${}^{101}$Ag decay & -10571.66\newline -28178.28\newline -35272.38\newline -44833.35\newline -60592.94\newline -75542.72\newline -15450.29 \\ \hline 
\textbf{${}^{100}$Pd\newline }$\varepsilonup$: 100~\%\textbf{} & 3.63 d\newline  & ~74.78\newline 84.00\newline 126.15\newline 158.87 & 48\newline 52\newline 7.8\newline 1.66\newline  & ${}^{102}$Pd(p,p2n)\newline ${}^{104}$Pd(p,p4n)\newline ${}^{105}$Pd(p,p5n)\newline ${}^{106}$Pd(p,p6n)\newline ${}^{108}$Pd(p,p8n)\newline ${}^{110}$Pd(p,p10n)\newline ${}^{100}$Ag decay & -18846.41\newline -36453.02\newline -43547.11\newline -53108.14\newline -68867.76\newline \newline -26718.02 \\ \hline 
\textbf{${}^{105}$Rh\newline }$\betaup$${}^{-}$: 100~\%\textbf{} & ~35.36 h~ & ~306.1\newline ~318.9\underbar{} & ~~5.1\newline ~19.1~ & ${}^{106}$Pd(p,2p)\newline ${}^{108}$Pd(p,2p2n)\newline ${}^{110}$Pd(p,2p4n) & -9345.82\newline -25105.42\newline -40055.23 \\ \hline 
\textbf{${}^{102m}$Rh\newline }140.7 keV\textbf{\newline }IT: 0.23 \%\newline ~$\varepsilonup$: 99.77 \%\newline \newline  & ~3.742 a & 631.29\newline 697.49\newline 766.84\newline 1046.59\newline 1112.84\underbar{} & 56.0\newline 44.0\newline 34.0\newline 34.0\newline 19.0 & ${}^{104}$Pd(p,2pn)\newline ${}^{105}$Pd(p,2p2n)\newline ${}^{106}$Pd(p,2p3n)\newline ${}^{108}$Pd(p,2p5n)\newline ${}^{110}$Pd(p,2p7n) & -17974.84\newline -25068.94\newline -34629.91\newline -50389.49\newline -65339.29 \\ \hline 
\textbf{${}^{102g}$Rh\newline }$\betaup$${}^{-}$: 22\textit{~}\%\newline EC: 78 \%~\newline \textbf{} & 207.3 d &   468.58 739.5  1158.10   &   2.9\newline  0.53   0.58 & ${}^{104}$Pd(p,2pn)\newline ${}^{105}$Pd(p,2p2n)\newline ${}^{106}$Pd(p,2p3n)\newline ${}^{108}$Pd(p,2p5n)\newline ${}^{110}$Pd(p,2p7n) & -17974.84\newline -25068.94\newline -34629.91\newline -50389.49\newline -65339.29 \\ \hline 
\textbf{${}^{101m}$Rh\newline }157.32 keV\textbf{\newline }IT: 7.20\%\textbf{\newline }EC: 92.8\%~\newline \textbf{} & 4.34 d~ & 306.857\newline ~545.117 & 81\newline 4.3 & ${}^{102}$Pd(p,2p)\newline ${}^{104}$Pd(p,2p2n)\newline ${}^{105}$Pd(p,2p3n)\newline ${}^{106}$Pd(p,2p4n)\newline ${}^{108}$Pd(p,2p6n)\newline ${}^{110}$Pd(p,2p8n)\newline ${}^{101}$Pd decay & -7809.14\newline -25415.76\newline -32509.86\newline -42070.83\newline -57830.41\newline -72780.20\newline -10571.66 \\ \hline 
\textbf{${}^{101g}$Rh\newline }$\varepsilonup$: 100~\%\textbf{\newline } & 3.3 a~ & 127.226\newline 198.01\newline 325.23\newline  & 68\newline 73\newline 11.8 & ${}^{102}$Pd(p,2p)\newline ${}^{104}$Pd(p,2p2n)\newline ${}^{105}$Pd(p,2p3n)\newline ${}^{106}$Pd(p,2p4n)\newline ${}^{108}$Pd(p,2p6n)\newline ${}^{110}$Pd(p,2p8n)\newline ${}^{101}$Pd decay & -7809.14\newline -25415.76\newline -32509.86\newline -42070.83\newline -57830.41\newline -72780.20\newline -10571.66 \\ \hline 
\textbf{${}^{100}$Rh\newline }EC: 96.1~\%~\newline $\betaup$${}^{+}$: 3.9 \%~\textbf{\newline } & 20.8 h\newline  & 446.153\newline 539.512\newline 822.654\newline 1107.223\newline 1553.348 & 11.98\newline 80.6\newline 21.09\newline 13.57\newline 20.67 & ${}^{102}$Pd(p,2pn)\newline ${}^{104}$Pd(p,2p3n)\newline ${}^{105}$Pd(p,2p4n)\newline ${}^{106}$Pd(p,2p5n)\newline ${}^{108}$Pd(p,2p7n)\newline ${}^{110}$Pd(p,2p9n) & -17703.1\newline -35309.7\newline -42403.8\newline -51964.7\newline -67724.3\newline - \\ \hline 
\end{tabular}

\end{center}
\end{table*}

\setcounter{table}{2}
\begin{table*}[t]
\tiny
\caption{continued}
\begin{center}
\begin{tabular}{|p{0.8in}|p{0.6in}|p{0.6in}|p{0.5in}|p{0.8in}|p{0.8in}|} \hline 
Nuclide\newline (isomeric state)\newline Decay path & Half-life & E${}_{?}$(keV) & I${}_{?}$(\%) & Contributing reaction & Q-value\newline (keV)\newline gs $\mathrm{\to}$ gs \\ \hline 

\textbf{$^{99m}$Rh\newline }64.3 keV\textbf{\newline }$\varepsilon$: $>$99.84 \%\newline IT:$<$0.16 \%\newline \textbf{} & 4.7 h & 340.8\newline 617.8\newline 1261.2 & 69\newline 11.8\newline ~10.9 & ${}^{102}$Pd(p,2p2n)\newline ${}^{104}$Pd(p,2p4n)\newline ${}^{105}$Pd(p,2p5n)\newline ${}^{106}$Pd(p,2p6n)\newline ${}^{108}$Pd(p,2p8n)\newline ${}^{110}$Pd(p,2p10n)\newline ${}^{99}$Pd decay & -25784.61\newline -43391.23\newline -50485.34\newline -60046.29\newline -75805.87\newline -\newline -29963.71 \\ \hline 
\textbf{$^{99g}$Rh\newline }EC: 96.4 \%\newline $\betaup^{+}$: 3.6 \%\newline \textbf{} & ~16.1 d & 89.76\newline 353.05\newline 528.24 & 33.4\newline 34.6\newline 38.0 & ${}^{102}$Pd(p,2p2n)\newline ${}^{104}$Pd(p,2p4n)\newline ${}^{105}$Pd(p,2p5n)\newline ${}^{106}$Pd(p,2p6n)\newline ${}^{108}$Pd(p,2p8n)\newline ${}^{110}$Pd(p,2p10n) & -25784.61\newline -43391.23\newline -50485.34\newline -60046.29\newline -75805.87\newline - \\ \hline 
\textbf{${}^{103}$Ru\newline }$\betaup$${}^{-}$: 100 \% & 39.247 d & 497.085 & 91.0  & ${}^{105}$Pd(p,3p)\newline ${}^{106}$Pd(p,3pn)\newline ${}^{108}$Pd(p,3p3n)\newline ${}^{110}$Pd(p,3p5n)\newline ${}^{103}$Tc decay & -15732.06\newline -25293.02\newline -41052.62\newline -56002.41\newline -27172.42 \\ \hline 
\textbf{${}^{97}$Ru\newline }$\varepsilonup$: 100 \%\textbf{} & 2.83 d & 215.70\newline 324.49 & 85.62\newline 10.79\newline  & ${}^{102}$Pd(p,3p3n)\newline ${}^{104}$Pd(p,3p5n)\newline ${}^{105}$Pd(p,3p6n)\newline ${}^{106}$Pd(p,3p7n)\newline ${}^{97}$Rh decay & -40603.59\newline -58210.2\newline -65304.29\newline -74865.26\newline -50481.75 \\ \hline 
\end{tabular}

\end{center}

\begin{flushleft}
\tiny{\noindent Increase the Q-values if compound particles are emitted by: np-d, +2.2 MeV; 2np-t, +8.48 MeV; n2p-${}^{3}$He, +7.72 MeV; 2n2p-$\alpha$, +28.30 MeV.
\noindent Decrease Q-values for isomeric states with level energy of the isomer
}
\end{flushleft}

\end{table*}

%\setcounter{table}{0}
%\begin{table*}[t]
%\tiny
%\caption{continued}
%\begin{center}

%\begin{flushleft}
%\tiny{\noindent 
%}
%\end{flushleft}

%\end{center}
%\end{table*}

\section{Results and discussion}
\label{3}

\subsection{Cross sections}
\label{3.1}

\noindent The measured cross sections for the production of $^{103,104m,104g,105g,106m,110m}$Ag, ${}^{100,101}$Pd, ${}^{99m,100mg,100,101m,101g,102m,102g,105}$Rh and $^{103,97}$Ru  are shown in Table 4-5 and Figures 1-18. The figures also show the earlier literature experimental results and the theoretical results calculated with the TALYS code \citep{Koning2012} as listed in the TENDL-2014 \citep{Koning2014} and 2015 \citep{Koning2015} on-line libraries. For total production cross sections also the ALICE-IPPE \citep{Dityuk} results are shown, taken from the MENDL-2P library \citep{Shubin} where these data are available for some assessed activation products.

\noindent Due to the experimental circumstances (stacked foil technique, large number of simultaneously irradiated target, high radiation dose rate at EOB, limited detector capacity, limited available time for spectra measurements) no cross section data were obtained for some short-lived activation products. In cases of low energy unresolved gamma-rays, small effective cross section or low abundance some characteristic $\gamma$-lines could not be identified in the measured spectra. If this happens to be the only $\gamma$-line of an activation product, unfortunately no cross-sections could be obtained.

\noindent The results are discussed separately for each reaction product. Naturally occurring palladium is composed of 6 stable isotopes (${}^{102}$Pd: 1.02 \%, $^{104}$Pd: 11.14 \%, $^{105}$Pd: 22.38 \%, $^{106}$Pd: 27.33 \%, $^{108}$Pd: 26.46 \% and $^{110}$Pd: 11.72 \%) \citep{Berglund}. The relevant contributing reactions are collected in Table. 3.

\noindent The silver isotopes are produced only through (p,xn) reactions, the palladium isotopes directly via (p,pxn) reactions and through the EC-$\beta^{+}$ decay of silver and ?${}^{- }$ decay of isobaric rhodium radio-parents, the rhodium radioisotopes are produced through direct (p,2pxn) reactions (including clustered particle emissions) and EC-$\beta^{+}$ decay of simultaneously produced isobaric palladium radio-products.

\noindent The results of our different irradiation series show only moderate agreement overall.  In case of low statistics, the disagreement is more or less understandable, but the systematic disagreements, mostly between values obtained from irradiations and measurements performed at another location, are more difficult to explain. The different experimental circumstances can result in deviations of the used beam intensities often due to different detector efficiencies that cannot be cross-checked between the measuring sites. Also monitors and targets were often measured at different detector-target distances, which could result in systematic offsets. In many cases automatic peak fitting was used with different commercial analyzing software, which could cause systematic errors. As in most cases no earlier experimental data were available, we thought it worthwhile to show all experimental results from our multiple datasets, even if in the name of scientific honesty, we recognize that some discrepancies exist.

\subsubsection{Radioisotopes of silver}
\label{3.1.1}

\vspace{.5 cm}
\noindent \textbf{\textit{3.1.1.1 ${}^{nat}$Pd(p,xn)${}^{110m}$Ag reaction}}

\noindent The radionuclide $^{110}$Ag has two long-lived states. The high spin isomeric state (I$^{\pi}$ = 6$^{+}$, T$_{1/2}$ = 249.8 d) decays with ${\beta}^{-}$ (98.67 \%) to ${}^{110}$Cd and with IT (1.33\%) to the short-lived ground state (I$^{\pi}$ = 1$^{+}$, T$_{1/2}$ = 24.6 s). Only the ${}^{110}$Pd(p,n) reaction is contributing. The measured and calculated cross-sections for ${}^{110m}$Ag production are shown in Fig. 1. All series of measurements are compatible with the published values below 10 MeV \citep{Batii, Dillmann2011}. The TENDL-2015 predictions describe well the region from threshold to maximum but overestimate above 12 MeV. The previous TENDL-2014 gives better prediction around the maximum. 

\begin{figure}
\includegraphics[width=0.5\textwidth]{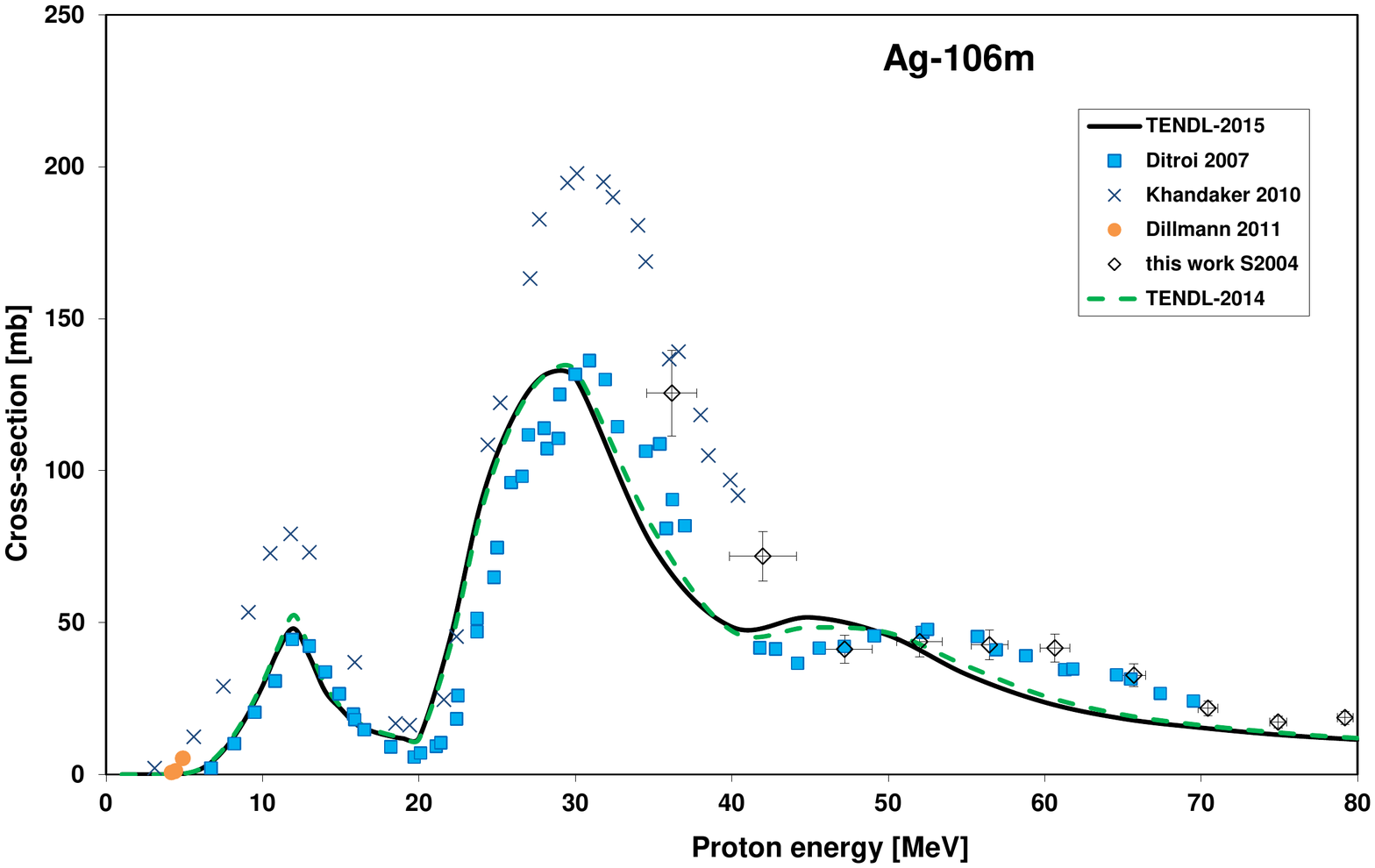}
\caption{Experimental and theoretical cross section values for proton induced ${}^{110m}$Ag production on natural palladium}
\label{fig:2}       
\end{figure}

\vspace{.5 cm}
\noindent \textbf{\textit{3.1.1.2}} \textbf{\textit{${}^{nat}$Pd(p,xn)${}^{106m}$Ag reaction }}

\noindent The production cross-sections for the long-lived isomeric state (I$^{\pi}$ =6${}^{+}$, T$_{1/2 }$= 8.28 d, $\varepsilonup$: 100~\%) were measured and are shown in Fig. 2. Our new values agree with the earlier experimental results of \citep{Khandaker2010} above 35 MeV, and agree well with the TENDL predictions. Also the agreement with the data of \citep{DF2007} is good while the few low energy point of \citep{Dillmann2011} are compatible with the data of \citep{Khandaker2010}.

\begin{figure}
\includegraphics[width=0.5\textwidth]{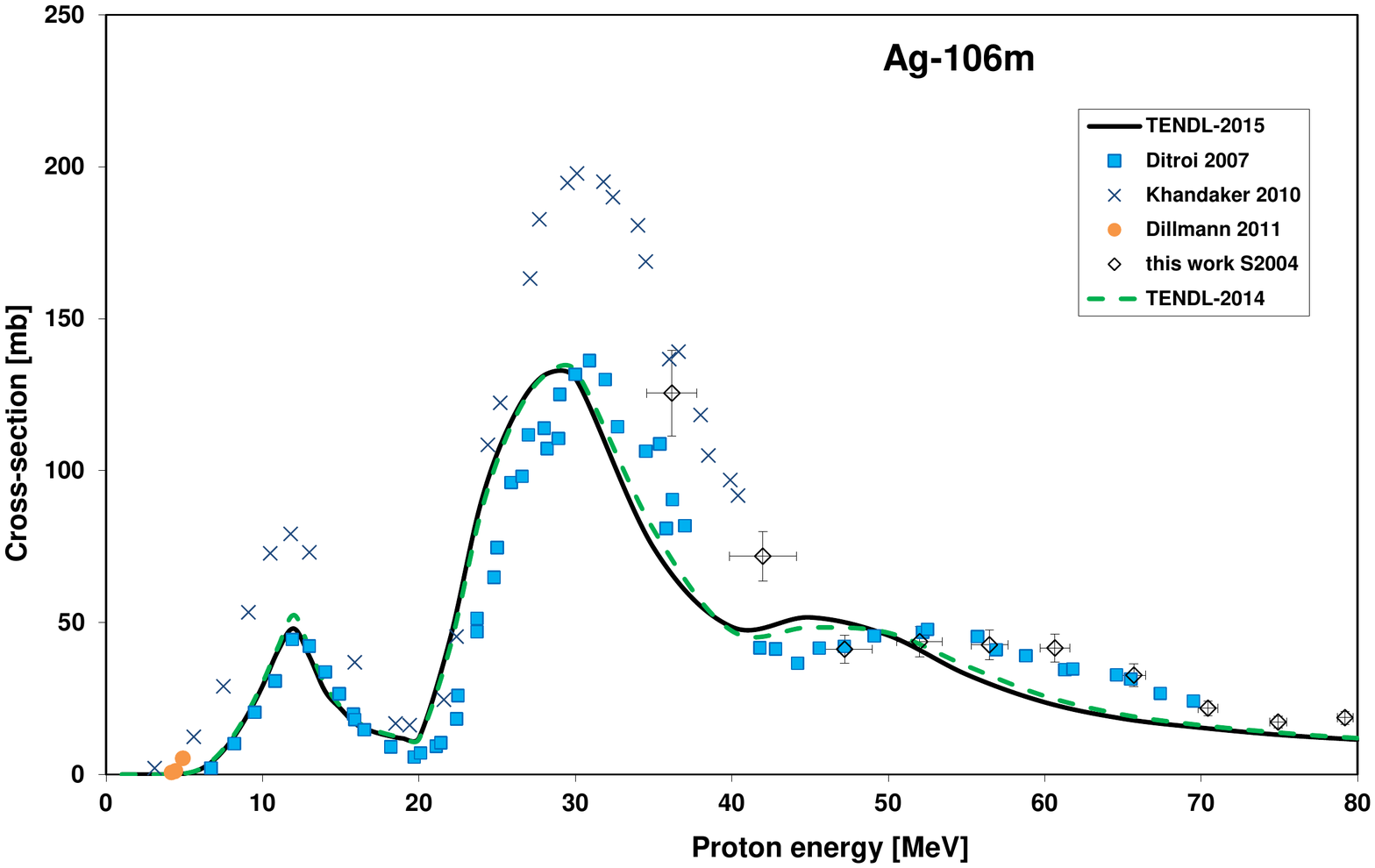}
\caption{Experimental and theoretical cross section values for proton induced ${}^{106m}$Ag production on natural palladium}
\label{fig:2}       
\end{figure}

\vspace{.5 cm}

\noindent \textbf{\textit{3.1.1.3 ${}^{nat}$Pd(p,xn)${}^{105m,g}$Ag reaction }}

\noindent \textbf{\textit{}}

\noindent The radionuclide ${}^{105}$Ag has also two long-lived states. The shorter-lived isomeric state (I${}^{?}$ = 7/2${}^{+}$, T${}_{1/2 }$= 7.7 min) decays with 99.66 \% by internal transition to the ground state (I$^{\pi}$ = $1/2^{-}$, T$_{1/2}$ = 41.3 d). The cumulative production cross sections for the ground state, after complete decay of the meta-stable state, were measured. The values are compared with the earlier results and with model calculation in Fig. 3. The agreement with the experimental data in the literature is acceptable both in shape and in the absolute values within 10 \%. The high energy part of the TENDL-2015 data seems to be shifted to lower energies.

\begin{figure}
\includegraphics[width=0.5\textwidth]{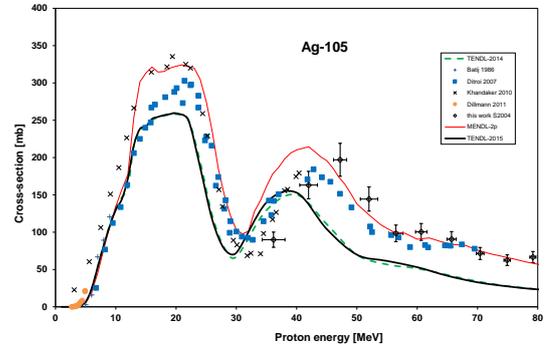}
\caption{Experimental and theoretical cross section values for proton induced ${}^{105}$Ag(cumulative) production on natural palladium}
\label{fig:3}       
\end{figure}

\vspace{.5 cm} 

\noindent \textbf{\textit{3.1.1.4 ${}^{nat}$Pd(p,xn)${}^{104m}$Ag and  ${}^{nat}$Pd(p,xn)${}^{104g}$Ag reactions }}

\noindent The radionuclide $^{104}$Ag has two states. The shorter-lived isomeric state (I$^{\pi}$ = 2$^{+}$, T$_{1/2 }$= 33.5 min) decays with 36.93 \% EC + 63 \% ${\beta}^{+}$ to ${}^{104}$Pd and by 0.07 \% internal transition to the ground state (I$^{pi}$ = 5$^{+}$, T$_{1/2 }$= 69.2 min). We could deduce independent production cross sections for both states after small corrections ($<$ 5\%) for internal transitions. The experimental and theoretical cross-sections are shown in Figs. 4-5. The agreement with the theory and the earlier experimental data is acceptable for the formation of $^{104m}$Ag although our data V2 do not represent properly the dip around 25 MeV and the TENDL-2015 values are shifted downward in energy. For formation of  ${}^{104g}$Ag  all datasets are agreeing well and show the contributions of the different stable target isotopes, except for the data above 13 MeV of \citep{Kormali}. For ${}^{104m}$Ag our new data do not agree with the earlier data sets in the overlapping energy region (only 1 data point), and those are even not compatible with each other. Because of the observable difference between TENDL-2014 and TENDL-2015, both curves are presented in Figs. 4 and 5.

\begin{figure}
\includegraphics[width=0.5\textwidth]{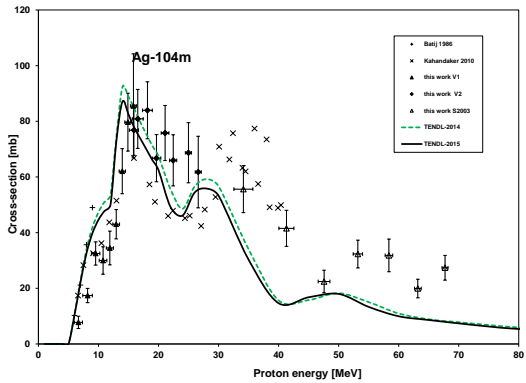}
\caption{Experimental and theoretical cross sections for proton induced ${}^{104m}$Ag production on natural palladium}
\label{fig:4}       
\end{figure}

\begin{figure}
\includegraphics[width=0.5\textwidth]{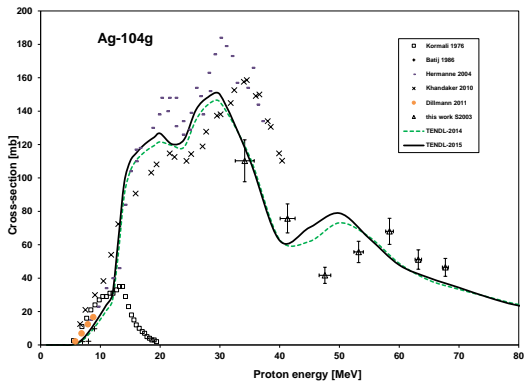}
\caption{Experimental and theoretical cross sections for proton induced ${}^{104g}$Ag production on natural palladium}
\label{fig:5}       
\end{figure}

\vspace{.5 cm} 

\noindent \textbf{\textit{3.1.1.5 $^{nat}$Pd(p,xn)${}^{103g}$Ag reaction }}

\noindent 

\noindent The experimental and theoretical activation cross-section data for $^{103}$Ag (I$^{\pi}$ = 7/2$^{+}$, T$_{1/2}$ = 65.7 min, $\varepsilonup$: 100 \%) are shown in Fig. 6. The production cross section of the ground state were deduced from spectra measured after complete decay of short-lived metastable state (I$^{\pi}$ = ${1/2}^{-}$ $T_{1/2}$ = 5.7 s, IT: 100 \%) and are hence cumulative. These particular data were already published in a special paper on $^{103}$Pd production \citep{TF2009}. The agreement with \citep{Hermanne2004} is good while the results of \citep{Khandaker2010} are somewhat lower and are not decreasing above 40 MeV. The both previous data sets are not compatible with each other. The TENDL values show clearly the multiple contributions of five stable Pd isotopes but seem to be shifted by 3-4 MeV.

\begin{figure}
\includegraphics[width=0.5\textwidth]{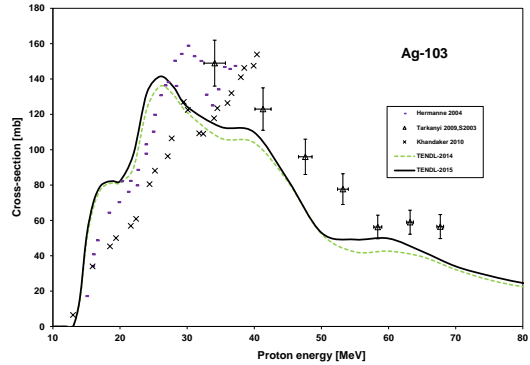}
\caption{Experimental and theoretical cross sections for proton induced $^{103}$Ag production on natural palladium}
\label{fig:6}       
\end{figure}
 
\subsubsection{Radioisotopes of palladium}
\label{3.1.2}

\vspace{.5 cm}
\noindent \textbf{\textit{3.1.2.1 $^{nat}$Pd(p,x)$^{101}$Pd  reaction}}

\noindent The $^{101}$Pd (I$^{\pi}$ = 5/2$^{+}$ T$_{1/2 }$= 8.47 h, EC: 92.8\%, $\beta^{+}$: 7.2 \%) is produced directly via (p,pxn) reaction and through the decay of the short-lived $^{101}$Ag parent ((I$^{\pi}$ = 9/2${}^{+}$, T$_{1/2 }$= 11.1 min, EC+ ${\beta}^{+}$: 100\%). The cumulative measured and calculated cross-sections are shown in Fig. 7. There is a good agreement with the earlier experimental data of \citep{Khandaker2010} and with the TALYS data in the TENDL-2015 library under 32 and above 55 MeV. An unexpected drop of some points of our V2 data set can be observed. MENDL-2P gives good estimation between 40 and 60 MeV.

\begin{figure}
\includegraphics[width=0.5\textwidth]{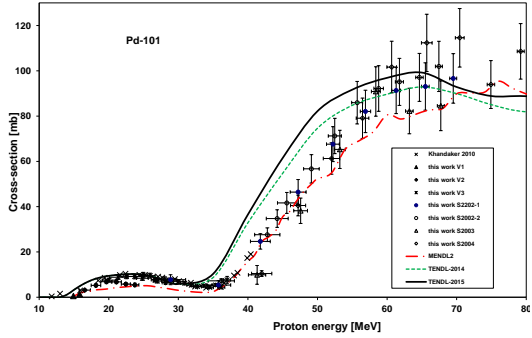}
\caption{Experimental and theoretical cross sections for proton induced $^{101}$Pd production on natural palladium}
\label{fig:7}       
\end{figure}

\vspace{.5 cm}

\noindent \textbf{\textit{3.1.2.2 $^{nat}$Pd(p,x)$^{100}$Pd reaction }}

\noindent The excitation function for production of ${}^{100}$Pd (I${}^{\pi}$ = 0$^{+}$, T$_{1/2}$= 3.63 d, $\varepsilon$: 100 \%), after the decay of both isomers of the short-lived $^{100m,g}$Ag (I$^{\pi}$ = 5$^{+}$, T$_{1/2 }$ = 2.01 min, EC+${\beta}^{+}$: 100 \% and  I$^{\pi}$ = 2$^{+}$, T$_{1/2 }$ = 2.24 min, EC+${\beta}^{+}$: 100 \%) parent nuclei is shown in Fig. 8. According to the figure the TENDL-2014 underestimates the experimental data. Below 40 MeV the data of \citep{Khandaker2010} are lower than our values in this energy region. A significant scattering can be observed in our V2 and V3 data sets under 30 MeV, probably because of the bad statistics. A significant improvement is seen in the TENDL-2015 version above 45 MeV. MENDL-2P shows a better trend above 60 MeV.

\begin{figure}
\includegraphics[width=0.5\textwidth]{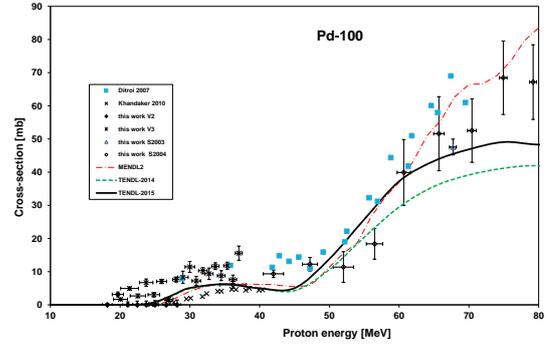}
\caption{Experimental and theoretical cross sections for proton induced $^{100}$Pd(cumulative)  production on natural palladium}
\label{fig:8}       
\end{figure}

\subsubsection{Radioisotopes of rhodium}
\label{3.1.3}

\vspace{.5 cm}
\noindent \textbf{\textit{3.1.3.1${}^{ }$${}^{nat}$Pd(p,x)${}^{1}$${}^{0}$${}^{5}$Rh reaction }}

\noindent The cumulative excitation functions for production of the $^{105g}$Rh ground state (I$^{\pi}$ = 7/2$^{+}$, T$_{1/2}$ = 35.36 h, ${\beta}^{-}$:100 \%), measured after complete isomeric decay of the short-lived metastable state$^{ 105m}$Rh (I$^{\pi}$ = ${1/2}^{-}$, T$_{1/2 }$= 42.9 s, IT: 100 \%) are shown in Fig. 9. The ${}^{105g}$Rh is produced via (p,2pxn) reactions on higher mass (106-110) stable Pd isotopes. The agreement with \citep{Khandaker2010} is very good but the theory significantly underestimates the experiment up to 40 MeV (Fig. 9). Our data show strong scattering and slight disagreement with each other above 40 MeV, probably because of the worse statistics, which is also reflected in the error bars. No significant difference was observed between the two TENDL versions.

\begin{figure}
\includegraphics[width=0.5\textwidth]{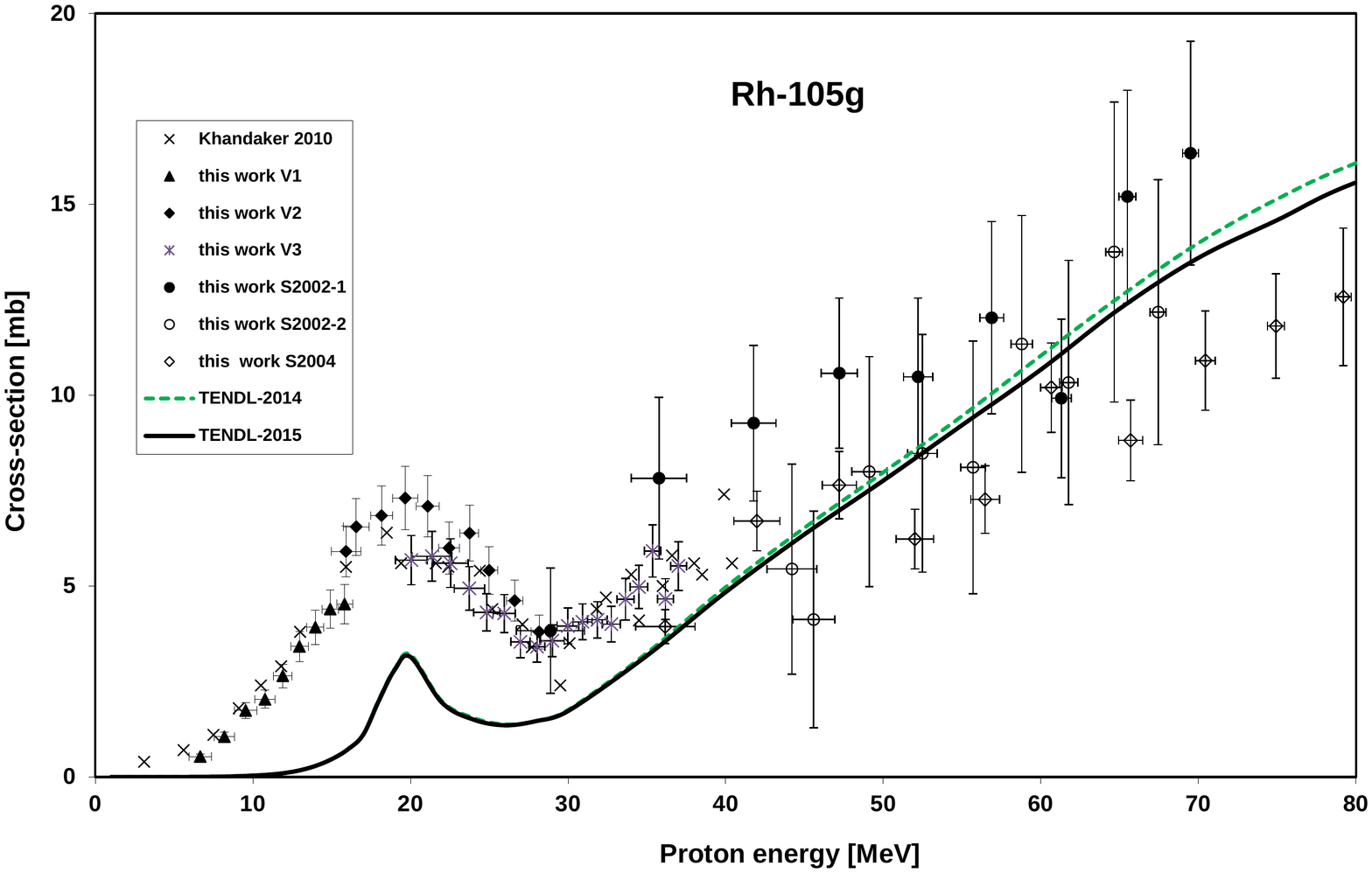}
\caption{Experimental and theoretical cross sections for proton induced $^{ 105}$Rh production on natural palladium}
\label{fig:9}       
\end{figure}

\vspace{.5 cm}
\noindent \textbf{\textit{3.1.3.2 $^{nat}$Pd(p,x)$^{102m}$Rh  and $^{nat}$Pd(p,x)$^{102g}$Rh  reactions }}

\noindent The radionuclide $^{102}$Rh has two long-lived states. The excitation functions for independent production of $^{102m}$Rh metastable state (I$^{\pi}$ = 0$^{+}$, T$_{1/2}$ = 3.742 a, EC: 99.767 \%, IT: 0.233 \%) and $^{102}$Rh ground state (I$^{\pi}$ = 1/2$^{-}$, T$_{1/2}$ = 207.3 d, EC: 62.3 \%, ${\beta}^{+}$: 15.7\%, ${\beta}^{-}$ 22 \%) are shown in Fig. 10 and 11. The data are strongly scattered because of the low statistics, but in average follow the trend of the TENDL curves for the $^{102m}$Rh isotope. The $^{102g}$Rh results are even more scattered and far above the TENDL-predictions.

\begin{figure}
\includegraphics[width=0.5\textwidth]{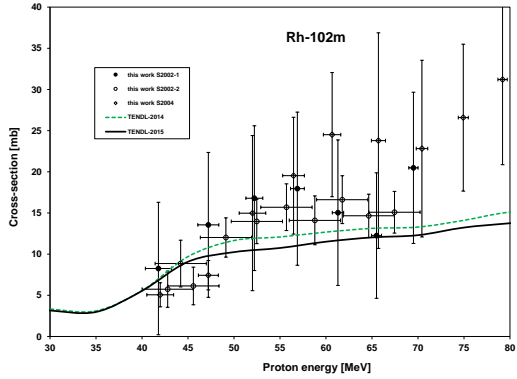}
\caption{Experimental and theoretical cross sections for proton induced $^{102m}$Rh production on natural palladium}
\label{fig:10}       
\end{figure}

\begin{figure}
\includegraphics[width=0.5\textwidth]{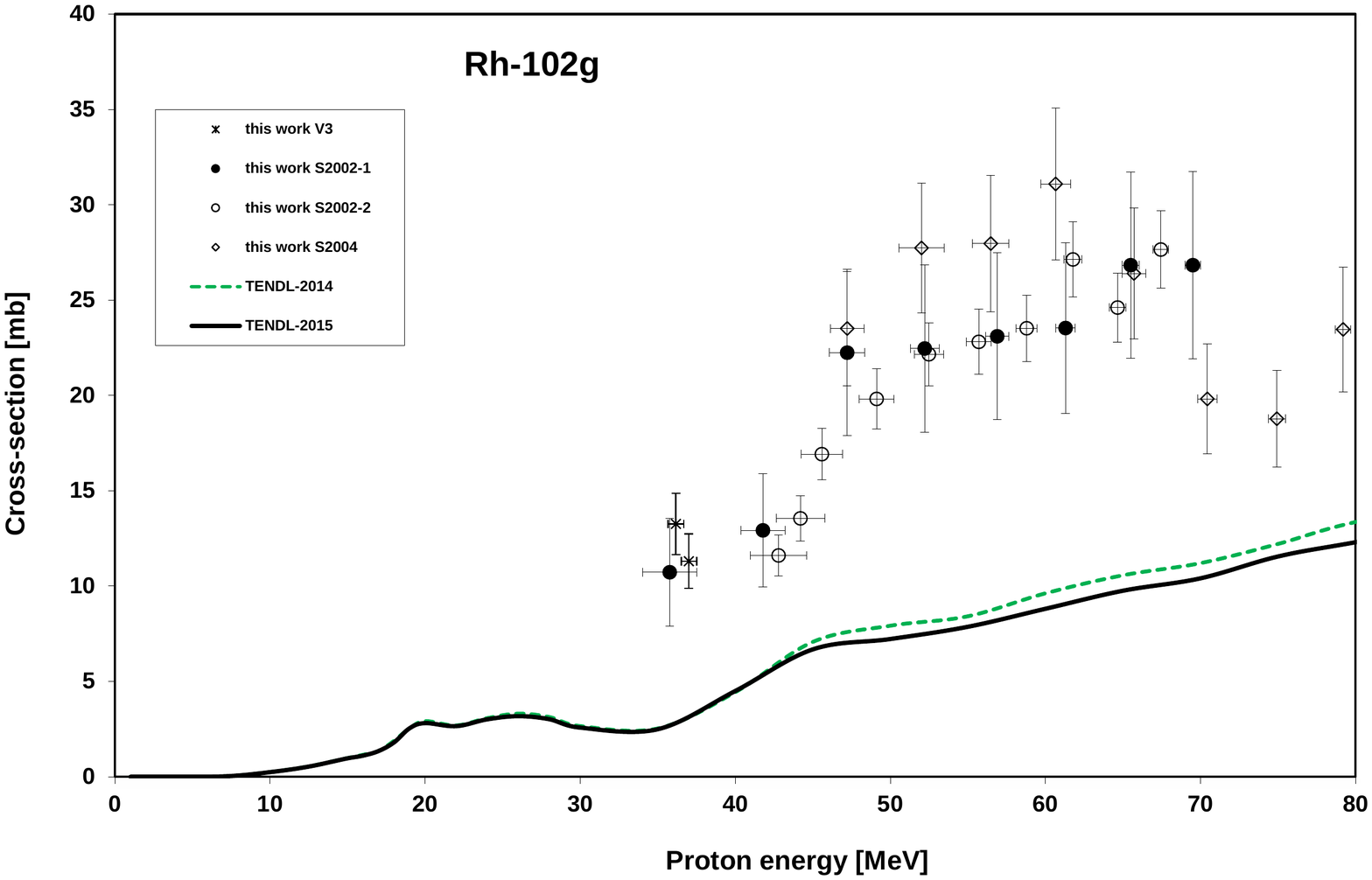}
\caption{Experimental and theoretical cross sections for proton induced $^{102g}$Rh production on natural palladium}
\label{fig:11}       
\end{figure}

\vspace{.5 cm}
\noindent \textbf{\textit{3.1.3.3 $^{nat}$Pd(p,x)$^{101m}$Rh  and $^{nat}$Pd(p,x)$^{101g}$Rh reactions }}

\noindent Out of the two longer-lived states of $^{101}$Rh, the $^{101m}$Rh isomeric state (I$^{\pi}$ = 9/2$^{+}$, T$_{1/2 }$= 4.34 d, IT: 7.2 \% EC: 92.8 \%) is produced directly through the $^{nat}$Pd(p,2pxn)$^{101m}$Rh process but is also fed from the mother $^{101}$Pd (T$_{1/2 }$ = 8.47 h).  

\noindent The cumulative cross sections of the $^{101m}$Rh isotope, deduced after complete decay of the $^{101}$Pd parent, is shown in Fig. 12. All data sets agree well and the theoretical prediction in TENDL-2015 only diverge above 55 MeV. There is a good agreement with the previous results and also with the TENDL predictions up to 60 MeV, above this energy the TENDL predictions slightly underestimate the experimental results.

\noindent The $^{101g}$Rh ground state (I$^{\pi}$ = ${1/2}^{-}$, T$_{1/2}$ = 3.3 a, $\varepsilon$: 100 \%) is produced directly and through the 7.2 \% internal decay of the meta-stable state.  The cumulative productions cross sections (m1+) obtained from the late spectra are shown in Fig. 13. The data points from our different experiments agree with each other, although the S2004 shows large scattering. The TENDL predictions underestimate the experimental results.

\begin{figure}
\includegraphics[width=0.5\textwidth]{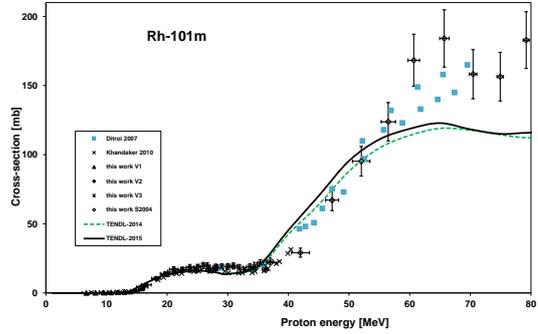}
\caption{Experimental and theoretical cross sections for proton induced $^{101m}$Rh production on natural palladium}
\label{fig:12}       
\end{figure}

\begin{figure}
\includegraphics[width=0.5\textwidth]{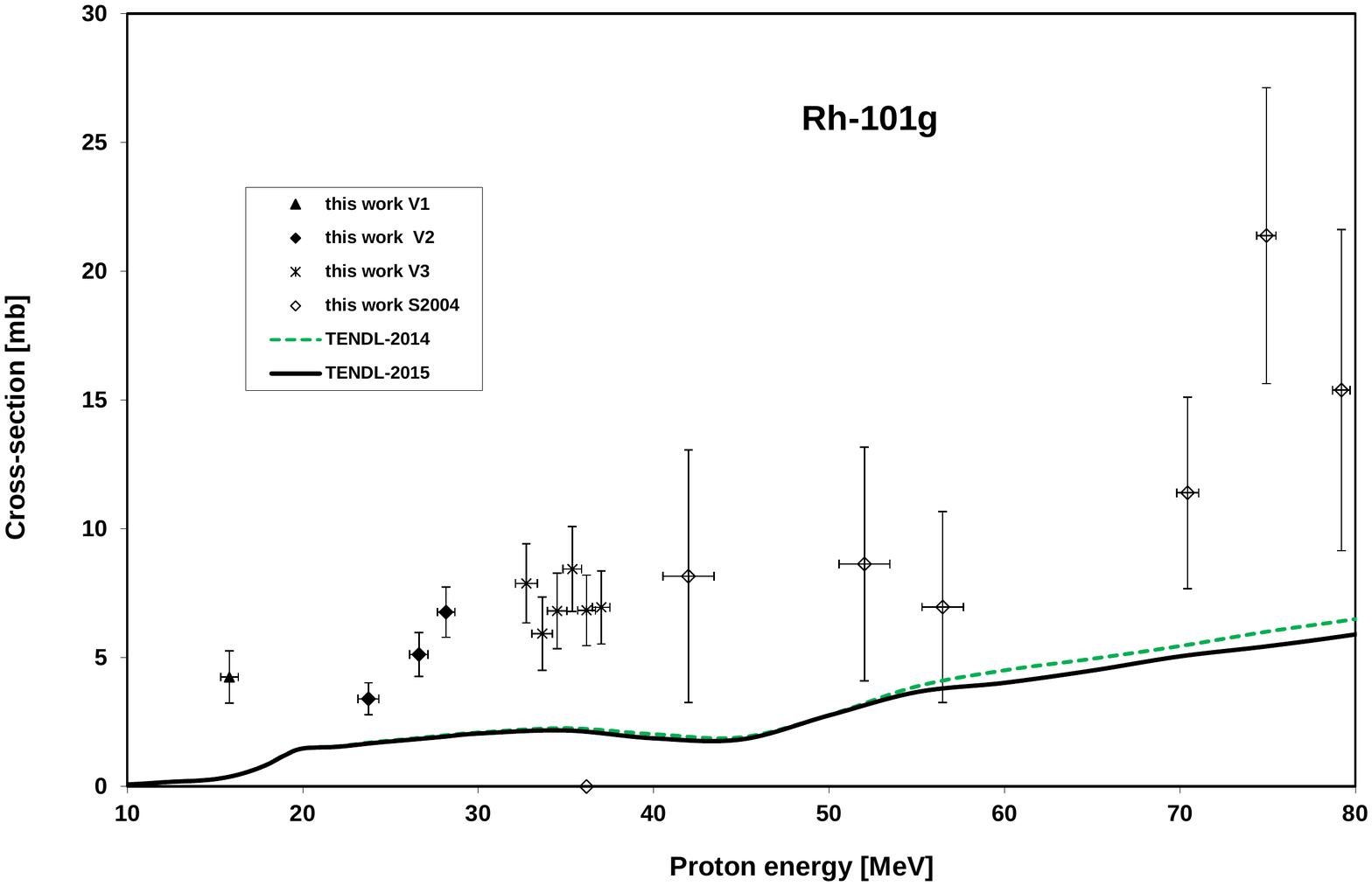}
\caption{Experimental and theoretical cross sections for proton induced $^{101g}$Rh(m+) production on natural palladium}
\label{fig:13}       
\end{figure}

\vspace{.5 cm}
\noindent \textbf{\textit{3.1.3.4 $^{nat}$Pd(p,x)$^{100g}$Rh reaction }}

\noindent The deduced cumulative cross sections of $^{100g}$Rh (I$^{\pi}$ =1$^{-}$, T$_{1/2}$ = 20.8 h, EC: 96.1 \%, ${\beta}^{+}$: 3.9 \%) (Fig. 14) include the direct production and the production through total decay of the short-lived isomeric state (I$^{\pi}$ = 5$^{+}$, T$_{1/2 }$= 4.6 min, IT 98.3: \%, EC: 1.7 \%). The contribution from the decay of $^{100}$Pd (3.63 d) was deduced. All experimental data-sets agree well and TENDL shows a larger dip than the experiments in the 40-60 MeV region. The MENDL-2P data represent only the direct production, that's why they show a stronger underestimation above 60 MeV.

\begin{figure}
\includegraphics[width=0.5\textwidth]{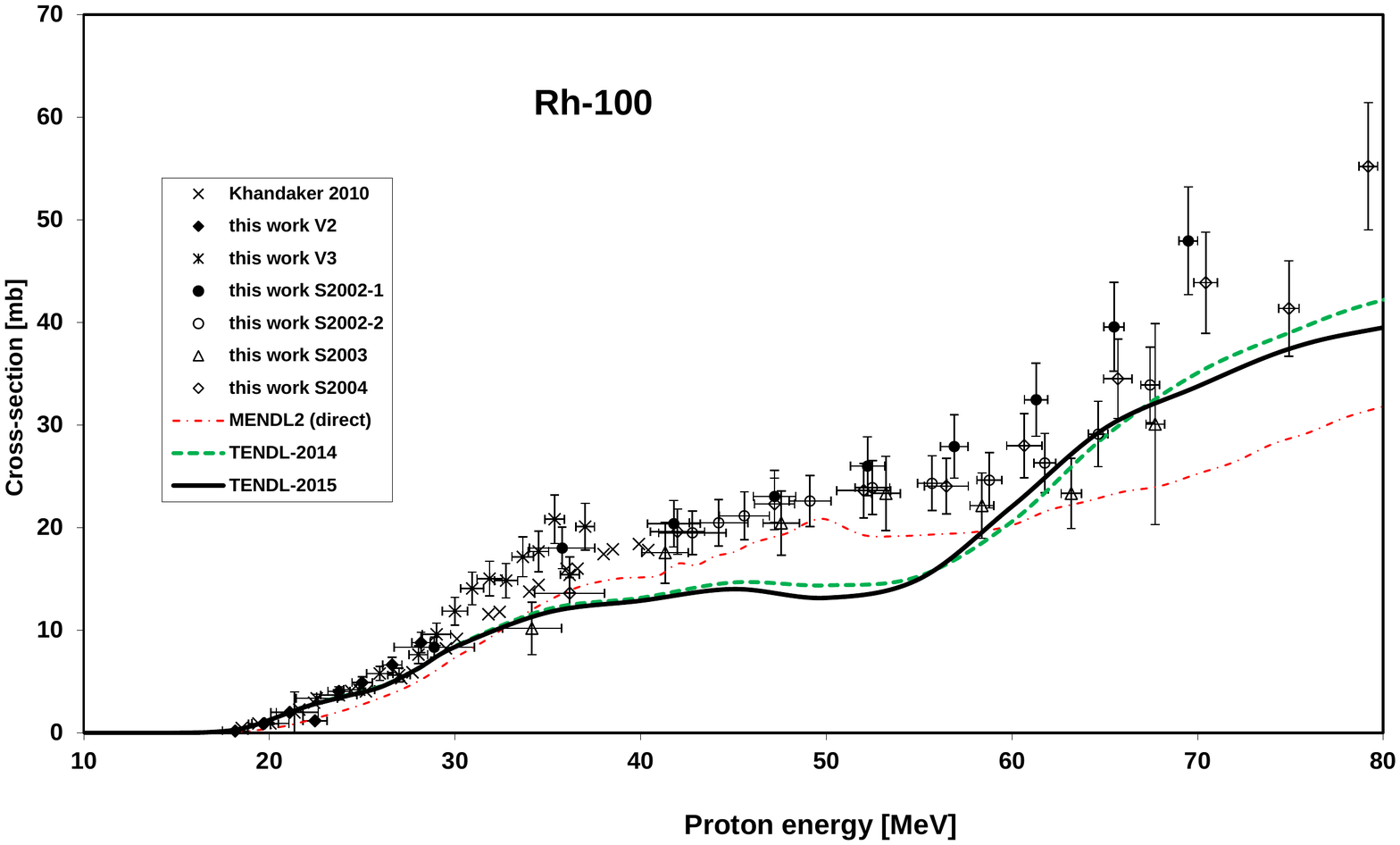}
\caption{Experimental and theoretical cross sections for proton induced $^{100}$Rh(cumulative) production on natural palladium}
\label{fig:14}       
\end{figure}

\vspace{.5 cm}
\noindent \textbf{\textit{3.1.3.5 $^{ }$$^{nat}$Pd(p,x)$^{99m}$Rh and $^{nat}$Pd(p,x)$^{99g}$Rh  reactions }}

\noindent The two long-lived states of $^{99}$Rh are decaying independently. They are produced directly and the metastable state $^{99m}$Rh is also populated through the decay of short-lived $^{99}$Pd parent (T$_{1/2 }$ = 21.4 min).  The excitation function for cumulative production of the $^{99m}$Rh metastable state (I$^{\pi}$ = 9/2$^{+}$, T$_{1/2}$= 4.7 h, IT $<$ 0.16 \%, EC: 92.7 \%, ${\beta}^{+}$: 7.3 \%), including the 97.4 \% decay contribution of $^{99}$Pd is shown in Fig. 15. The agreement between our different data sets is acceptable good in the overlapping energy region. The data set V3 had a much better statistic, as seen in the error bars. The TENDL calculations give only good estimation up to 37 MeV, above this energy they underestimate up to 70 MeV. MENDL-2P gives a strange curve, which does not follow the trend of the experiment at all. 

\noindent The cumulative cross sections of the $^{99g}$Rh ground state (I$^{\pi}$ = ${1/2}^-$ T$_{1/2}$ = 16.1 d, EC: 96.4 \%, ${\beta}^{+}$: 3.6 \%), including 2.6 \% decay contribution from $^{99}$Pd parent are shown in Fig. 16. There is an acceptable agreement between the different measurements series. The trend and magnitude of the TENDL predictions follow well the experimental results up to 60 MeV, but above this value the new TENDL-2015 gives overestimated results. It is difficult to decide, which TENDL version gives better estimation.

\begin{figure}
\includegraphics[width=0.5\textwidth]{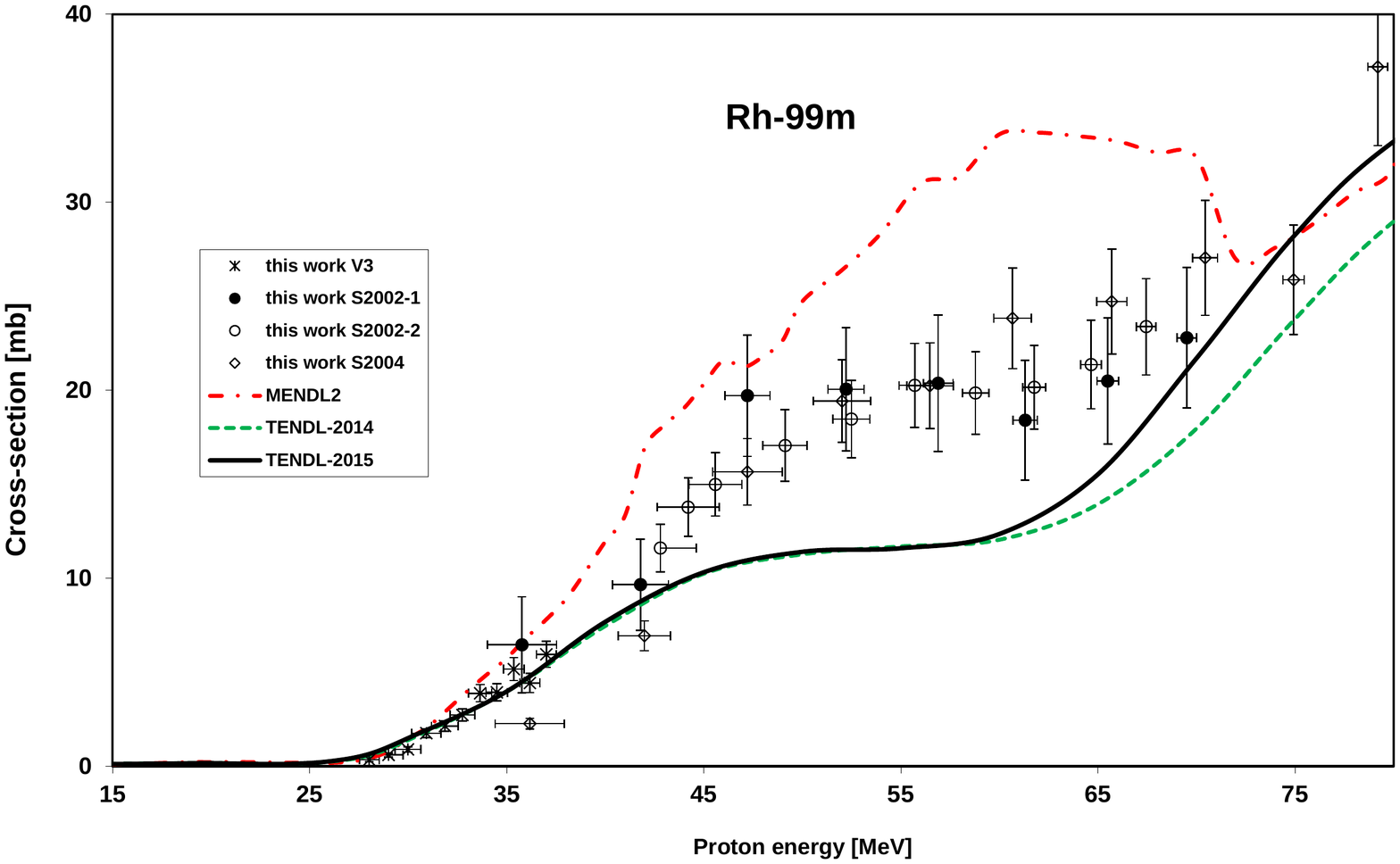}
\caption{Experimental and theoretical cross sections for proton induced $^{99m}$Rh production on natural palladium}
\label{fig:15}       
\end{figure}

\begin{figure}
\includegraphics[width=0.5\textwidth]{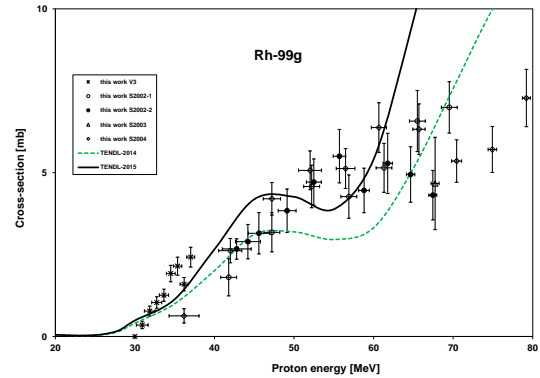}
\caption{Experimental and theoretical cross sections for proton induced $^{99g}$Rh production on natural palladium}
\label{fig:16}       
\end{figure}

\subsubsection{Radioisotopes of ruthenium}
\label{3.1.4}

\vspace{.5 cm}
\noindent \textbf{\textit{3.1.4.1 $^{nat}$Pd(p,x)$^{103}$Ru  reaction }}

\noindent The cumulative cross sections of the $^{103}$Ru ground state (I$^{\pi}$ = ${3/2}^{+}$ T$_{1/2}$ = 39.247 d, $\betaup^{-}$: 100 \%), including 100 \% decay contribution from $^{103}$Tc parent (T$_{1/2}$ = 54.2 s) are shown in Fig. 17. The TENDL predictions are significantly lower compared to the experiment. The different data sets are in acceptable agreement with each other, in spite of the scattering.

\begin{figure}
\includegraphics[width=0.5\textwidth]{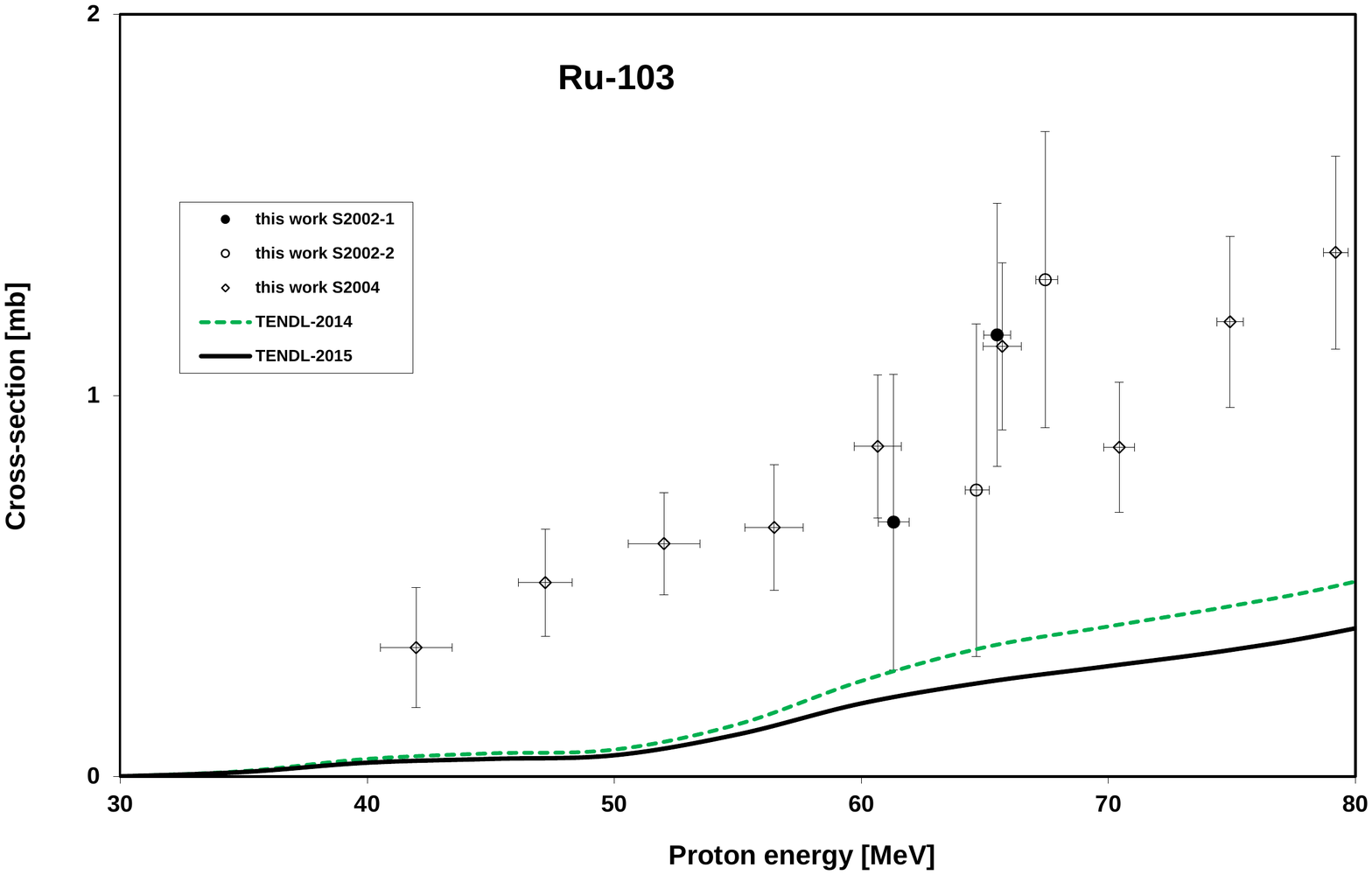}
\caption{Experimental and theoretical cross sections for the proton induced $^{103}$Ru production on natural palladium}
\label{fig:17}       
\end{figure}

\vspace{.5 cm}
\noindent \textbf{\textit{3.1.4.2 $^{nat}$Pd(p,xn)$^{97}$Ru reaction }}

\noindent The measured experimental cross section data of $^{97}$Ru (I$^{\pi}$ = ${5/2}^{+}$, T$_{1/2}$ = 2.83 d, $\varepsilon$: 100 \%) are cumulative (Fig. 18) including the direct reaction and indirect formation through the $^{97}$Ag (T$ _{1/2}$ = 25.5 s) $\longrightarrow ^{97}$Pd (T$_{1/2}$ = 3.1 min) $\longrightarrow ^{97m,g}$Rh (T$_{1/2}$ = 30.7 min and 46.2 min) $\longrightarrow ^{97}$Ru decay chain. All measurements agree well, except some single points. Both TENDL versions underestimate the experimental values above 65 MeV.

\begin{figure}
\includegraphics[width=0.5\textwidth]{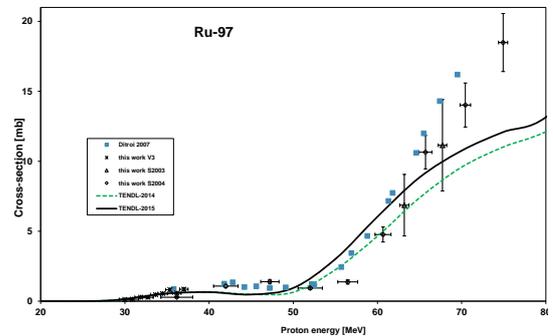}
\caption{Experimental and theoretical cross sections for proton induced $^{97}$Ru(cumulative) production on natural palladium}
\label{fig:18}       
\end{figure}

\begin{table*}[t]
\tiny
\caption{Experimental cross sections of $^{nat}$Pd(p,xn)$^{103,104m,104g,105g,106m,110m}$Ag, $^{100,101}$Pd and $^{97}$Ru reactions}
\begin{center}
\begin{tabular}{|p{0.1in}|p{0.1in}|p{0.1in}|p{0.2in}|p{0.1in}|p{0.2in}|p{0.1in}|p{0.2in}|p{0.1in}|p{0.2in}|p{0.1in}|p{0.2in}|p{0.1in}|p{0.2in}|p{0.1in}|p{0.2in}|p{0.1in}|p{0.2in}|p{0.1in}|p{0.2in}|p{0.1in}|} \hline 
\multicolumn{2}{|c|}{\textbf{E$\pm\Delta$}E (MeV)} & \textbf{Exp.} & \multicolumn{18}{|c|}{\textbf{$\sigma\pm\Delta\sigma$} (mb)} \\ \hline 
\multicolumn{2}{|c|}{\textbf{}} & \textbf{} & \multicolumn{2}{|c|}{\textbf{$^{110m}$Ag}} & \multicolumn{2}{|c|}{\textbf{$^{106m}$Ag}} & \multicolumn{2}{|c|}{\textbf{$^{105}$Ag}} & \multicolumn{2}{|c|}{\textbf{$^{104m}$Ag}} & \multicolumn{2}{|c|}{\textbf{$^{104g}$Ag}} & \multicolumn{2}{|c|}{\textbf{$^{103}$Ag}} & \multicolumn{2}{|c|}{\textbf{$^{101}$Pd}} & \multicolumn{2}{|c|}{\textbf{$^{100}$Pd}} & \multicolumn{2}{|c|}{\textbf{$^{97}$Ru}} \\ \hline 
15.8 & 0.5 & V1 & 1.81 & 0.32 & ~ &  & ~ &  & 85.53 & 18.69 & ~ &  & ~ &  & 1.41 & 0.17 & ~ &  & ~ &  \\ \hline 
14.9 & 0.5 &  & ~ &  & ~ &  & ~ &  & 79.70 & 10.40 & ~ &  & ~ &  & 0.55 & 0.08 & ~ &  & ~ &  \\ \hline 
14.0 & 0.5 &  & 3.32 & 0.42 & ~ &  & ~ &  & 62.07 & 8.10 & ~ &  & ~ &  & ~ &  & ~ &  & ~ &  \\ \hline 
13.0 & 0.6 &  & 4.43 & 0.59 & ~ &  & ~ &  & 43.02 & 5.27 & ~ &  & ~ &  & ~ &  & ~ &  & ~ &  \\ \hline 
11.9 & 0.6 &  & 5.84 & 0.70 & ~ &  & ~ &  & 34.48 & 6.09 & ~ &  & ~ &  & ~ &  & ~ &  & ~ &  \\ \hline 
10.8 & 0.6 &  & 7.06 & 0.82 & ~ &  & ~ &  & 29.99 & 4.92 & ~ &  & ~ &  & ~ &  & ~ &  & ~ &  \\ \hline 
9.5 & 0.7 &  & 8.91 & 1.02 & ~ &  & ~ &  & 32.51 & 4.15 & ~ &  & ~ &  & ~ &  & ~ &  & ~ &  \\ \hline 
8.2 & 0.8 &  & 6.46 & 0.75 & ~ &  & ~ &  & 17.40 & 2.57 & ~ &  & ~ &  & ~ &  & ~ &  & ~ &  \\ \hline 
6.7 & 0.9 & ~ & ~ & ~ & ~ & ~ & ~ & ~ & 7.78 & 2.20 & ~ & ~ & ~ & ~ & ~ & ~ & ~ & ~ & ~ & ~ \\ \hline 
28.2 & 0.5 & V2 & 1.66 & 0.31 & ~ &  & ~ &  & ~ &  & ~ &  & ~ &  & 6.48 & 0.84 & 4.75 & 0.54 & ~ &  \\ \hline 
26.6 & 0.5 &  & 1.68 & 0.30 & ~ &  & ~ &  & 61.76 & 12.87 & ~ &  & ~ &  & 7.78 & 1.01 & 7.55 & 0.85 & ~ &  \\ \hline 
25.0 & 0.5 &  & 1.67 & 0.29 & ~ &  & ~ &  & 68.69 & 10.80 & ~ &  & ~ &  & 9.79 & 1.20 & 8.42 & 0.95 & ~ &  \\ \hline 
23.7 & 0.6 &  & 1.67 & 0.26 & ~ &  & ~ &  & ~ &  & ~ &  & ~ &  & 5.46 & 0.76 & 3.59 & 0.41 & ~ &  \\ \hline 
22.4 & 0.6 &  & ~ &  & ~ &  & ~ &  & 65.96 & 9.19 & ~ &  & ~ &  & 5.84 & 0.66 & 6.65 & 0.77 & ~ &  \\ \hline 
21.1 & 0.7 &  & 1.30 & 0.23 & ~ &  & ~ &  & 75.79 & 9.90 & ~ &  & ~ &  & 6.79 & 0.77 & 5.17 & 0.58 & ~ &  \\ \hline 
19.7 & 0.8 &  & ~ &  & ~ &  & ~ &  & 66.73 & 8.47 & ~ &  & ~ &  & 6.88 & 0.86 & 3.14 & 0.36 & ~ &  \\ \hline 
18.2 & 0.9 &  & 1.55 & 0.20 & ~ &  & ~ &  & 83.95 & 10.23 & ~ &  & ~ &  & 5.26 & 0.66 & 0.21 & 0.06 & ~ &  \\ \hline 
16.5 & 1.0 &  & 1.48 & 0.20 & ~ &  & ~ &  & 80.87 & 10.60 & ~ &  & ~ &  & 3.15 & 0.44 &  &  & ~ &  \\ \hline 
15.9 & 0.8 & ~ & ~ & ~ & ~ & ~ & ~ & ~ & 76.81 & 9.03 & ~ & ~ & ~ & ~ & ~ & ~ & ~ & ~ & ~ & ~ \\ \hline 
37.0 & 0.5 & V3 & 1.07 & 0.31 & ~ &  & ~ &  & ~ &  & ~ &  & ~ &  & 7.69 & 1.51 & 15.58 & 2.15 & 0.85 & 0.10 \\ \hline 
36.2 & 0.5 &  & ~ &  & ~ &  & ~ &  & ~ &  & ~ &  & ~ &  & 4.45 & 0.52 & 7.62 & 1.35 & 0.57 & 0.07 \\ \hline 
35.4 & 0.5 &  & ~ &  & ~ &  & ~ &  & ~ &  & ~ &  & ~ &  & 5.41 & 0.64 & 11.80 & 1.06 & 0.81 & 0.10 \\ \hline 
34.5 & 0.5 &  & ~ &  & ~ &  & ~ &  & ~ &  & ~ &  & ~ &  & 4.41 & 0.51 & 8.78 & 1.50 & 0.55 & 0.07 \\ \hline 
33.6 & 0.6 &  & ~ &  & ~ &  & ~ &  & ~ &  & ~ &  & ~ &  & 5.09 & 0.59 & 11.64 & 1.02 & 0.46 & 0.06 \\ \hline 
32.7 & 0.6 &  & ~ &  & ~ &  & ~ &  & ~ &  & ~ &  & ~ &  & 4.75 & 0.55 & 9.36 & 1.73 & 0.28 & 0.04 \\ \hline 
31.9 & 0.7 &  & ~ &  & ~ &  & ~ &  & ~ &  & ~ &  & ~ &  & 5.32 & 0.61 & 10.29 & 0.94 & 0.28 & 0.04 \\ \hline 
30.9 & 0.7 &  & ~ &  & ~ &  & ~ &  & ~ &  & ~ &  & ~ &  & 6.18 & 0.71 & 7.22 & 1.37 & 0.14 & 0.03 \\ \hline 
30.0 & 0.8 &  & 1.22 & 0.31 & ~ &  & ~ &  & ~ &  & ~ &  & ~ &  & 6.91 & 0.79 & 11.44 & 1.66 & 0.12 & 0.03 \\ \hline 
29.0 & 0.7 &  & ~ &  & ~ &  & ~ &  & ~ &  & ~ &  & ~ &  & 7.53 & 0.86 & 8.33 & 1.82 & ~ &  \\ \hline 
28.0 & 0.5 &  & ~ &  & ~ &  & ~ &  & ~ &  & ~ &  & ~ &  & 8.07 & 0.92 & 7.62 & 0.77 & ~ &  \\ \hline 
27.0 & 0.6 &  & ~ &  & ~ &  & ~ &  & ~ &  & ~ &  & ~ &  & 7.13 & 0.84 & 1.69 & 0.61 & ~ &  \\ \hline 
25.9 & 0.7 &  & 1.39 & 0.31 & ~ &  & ~ &  & ~ &  & ~ &  & ~ &  & 9.55 & 1.09 & 7.05 & 0.65 & ~ &  \\ \hline 
24.8 & 0.8 &  & 1.53 & 0.20 & ~ &  & ~ &  & ~ &  & ~ &  & ~ &  & 8.40 & 0.95 & 3.04 & 0.66 & ~ &  \\ \hline 
23.7 & 1.0 &  & 1.05 & 0.25 & ~ &  & ~ &  & ~ &  & ~ &  & ~ &  & 9.15 & 1.04 & 6.71 & 1.10 & ~ &  \\ \hline 
22.5 & 1.1 &  & 1.58 & 0.19 & ~ &  & ~ &  & ~ &  & ~ &  & ~ &  & 9.22 & 1.05 & 2.69 & 0.55 & ~ &  \\ \hline 
21.4 & 1.3 &  & 1.58 & 0.19 & ~ &  & ~ &  & ~ &  & ~ &  & ~ &  & 8.85 & 1.00 & 4.94 & 0.43 & ~ &  \\ \hline 
20.1 & 1.4 & ~ & 1.50 & 0.18 & ~ & ~ & ~ & ~ & ~ & ~ & ~ & ~ & ~ & ~ & 7.58 & 0.86 & 1.65 & 0.32 & ~ & ~ \\ \hline 
79.2 & 0.5 & S2004 & ~ &  & 18.68 & 2.11 & 66.67 & 7.52 & ~ &  & ~ &  & ~ &  & 108.60 & 12.25 & 67.17 & 11.27 & 25.83 & 2.90 \\ \hline 
74.9 & 0.5 &  & ~ &  & 17.29 & 1.95 & 62.51 & 7.05 & ~ &  & ~ &  & ~ &  & 93.93 & 10.58 & 68.45 & 11.06 & 18.49 & 2.08 \\ \hline 
70.4 & 0.6 &  & ~ &  & 21.78 & 2.46 & 71.52 & 8.07 & ~ &  & ~ &  & ~ &  & 114.58 & 12.91 & 52.54 & 9.56 & 14.02 & 1.58 \\ \hline 
65.7 & 0.8 &  & ~ &  & 32.70 & 3.69 & 90.62 & 10.21 & ~ &  & ~ &  & ~ &  & 112.34 & 12.66 & 51.60 & 11.17 & 10.65 & 1.20 \\ \hline 
60.7 & 1.0 &  & ~ &  & 41.61 & 4.68 & 100.38 & 11.30 & ~ &  & ~ &  & ~ &  & 101.66 & 11.42 & 39.92 & 9.92 & 4.77 & 0.54 \\ \hline 
56.5 & 1.2 &  & ~ &  & 42.67 & 4.80 & 98.39 & 11.08 & ~ &  & ~ &  & ~ &  & 79.06 & 8.92 & 18.38 & 4.63 & 1.38 & 0.16 \\ \hline 
52.0 & 1.4 &  & ~ &  & 43.72 & 4.92 & 144.36 & 16.22 & ~ &  & ~ &  & ~ &  & 61.28 & 6.92 & 11.39 & 4.61 & 0.93 & 0.11 \\ \hline 
47.2 & 1.1 &  & ~ &  & 41.22 & 4.64 & 196.98 & 22.13 & ~ &  & ~ &  & ~ &  & 40.52 & 4.59 & 12.25 & 2.04 & 1.39 & 0.16 \\ \hline 
42.0 & 1.5 &  & ~ &  & 71.81 & 8.07 & 163.15 & 18.34 & ~ &  & ~ &  & ~ &  & 10.45 & 1.26 & 9.34 & 1.12 & 1.08 & 0.12 \\ \hline 
36.2 & 1.9 & ~ & ~ & ~ & 125.50 & 14.10 & 89.97 & 10.13 & ~ & ~ & ~ & ~ & ~ & ~ & 7.23 & 0.89 & 6.02 & 1.00 & 0.29 & 0.04 \\ \hline 
67.7 & 0.5 & S2003 & ~ &  & ~ &  & ~ &  & 27.38 & 4.44 & 46.48 & 5.36 & 56.5 & 6.8 & 84.56 & 11.00 & 47.57 & 2.38 & 11.15 & 3.27 \\ \hline 
63.2 & 0.5 &  & ~ &  & ~ &  & ~ &  & 19.92 & 3.36 & 51.15 & 5.78 & 59 & 6.8 & 82.24 & 10.07 & ~ &  & 6.86 & 2.20 \\ \hline 
58.4 & 0.6 &  & ~ &  & ~ &  & ~ &  & 31.83 & 5.89 & 68.05 & 7.81 & 56.3 & 6.7 & 90.87 & 10.91 & ~ &  & ~ &  \\ \hline 
53.2 & 0.8 &  & ~ &  & ~ &  & ~ &  & 32.31 & 5.01 & 55.70 & 6.41 & 77.7 & 8.7 & 65.31 & 8.44 & ~ &  & ~ &  \\ \hline 
47.6 & 1.0 &  & ~ &  & ~ &  & ~ &  & 22.45 & 4.07 & 41.71 & 4.84 & 96 & 10 & 38.20 & 5.61 & ~ &  & ~ &  \\ \hline 
41.3 & 1.2 &  & ~ &  & ~ &  & ~ &  & 41.56 & 6.47 & 75.73 & 8.68 & 123 & 12 & 9.88 & 4.14 & ~ &  & ~ &  \\ \hline 
34.1 & 1.6 & ~ & ~ & ~ & ~ & ~ & ~ & ~ & 55.62 & 8.45 & 110.26 & 12.56 & 149 & 13 & ~ & ~ & ~ & ~ & ~ & ~ \\ \hline 
69.5 & 0.5 & S2002-1 & ~ &  & ~ &  & ~ &  & ~ &  & ~ &  & ~ &  & 96.65 & 10.92 & ~ &  & ~ &  \\ \hline 
65.5 & 0.5 &  & ~ &  & ~ &  & ~ &  & ~ &  & ~ &  & ~ &  & 93.07 & 10.45 & ~ &  & ~ &  \\ \hline 
61.3 & 0.6 &  & ~ &  & ~ &  & ~ &  & ~ &  & ~ &  & ~ &  & 91.33 & 10.23 & ~ &  & ~ &  \\ \hline 
56.9 & 0.7 &  & ~ &  & ~ &  & ~ &  & ~ &  & ~ &  & ~ &  & 82.07 & 9.37 & ~ &  & ~ &  \\ \hline 
52.2 & 0.9 &  & ~ &  & ~ &  & ~ &  & ~ &  & ~ &  & ~ &  & 67.61 & 7.71 & ~ &  & ~ &  \\ \hline 
47.2 & 1.1 &  & ~ &  & ~ &  & ~ &  & ~ &  & ~ &  & ~ &  & 46.46 & 5.52 & ~ &  & ~ &  \\ \hline 
41.8 & 1.4 &  & ~ &  & ~ &  & ~ &  & ~ &  & ~ &  & ~ &  & 24.67 & 3.38 & ~ &  & ~ &  \\ \hline 
35.8 & 1.8 &  & ~ &  & ~ &  & ~ &  & ~ &  & ~ &  & ~ &  & 5.39 & 1.89 & ~ &  & ~ &  \\ \hline 
28.9 & 2.2 & ~ & ~ & ~ & ~ & ~ & ~ & ~ & ~ & ~ & ~ & ~ & ~ & ~ & 7.72 & 2.19 & ~ & ~ & ~ & ~ \\ \hline 
67.4 & 0.5 & S2002-2 & ~ &  & ~ &  & ~ &  & ~ &  & ~ &  & ~ &  & 101.94 & 11.09 & ~ &  & ~ &  \\ \hline 
64.6 & 0.5 &  & ~ &  & ~ &  & ~ &  & ~ &  & ~ &  & ~ &  & 97.10 & 10.59 & ~ &  & ~ &  \\ \hline 
61.8 & 0.6 &  & ~ &  & ~ &  & ~ &  & ~ &  & ~ &  & ~ &  & 95.12 & 10.39 & ~ &  & ~ &  \\ \hline 
58.8 & 0.7 &  & ~ &  & ~ &  & ~ &  & ~ &  & ~ &  & ~ &  & 92.19 & 10.07 & ~ &  & ~ &  \\ \hline 
55.7 & 0.8 &  & ~ &  & ~ &  & ~ &  & ~ &  & ~ &  & ~ &  & 85.91 & 9.40 & ~ &  & ~ &  \\ \hline 
52.5 & 0.9 &  & ~ &  & ~ &  & ~ &  & ~ &  & ~ &  & ~ &  & 71.21 & 7.81 & ~ &  & ~ &  \\ \hline 
49.1 & 1.1 &  & ~ &  & ~ &  & ~ &  & ~ &  & ~ &  & ~ &  & 56.71 & 6.26 & ~ &  & ~ &  \\ \hline 
45.6 & 1.3 &  & ~ &  & ~ &  & ~ &  & ~ &  & ~ &  & ~ &  & 41.65 & 4.63 & ~ &  & ~ &  \\ \hline 
44.2 & 1.6 &  & ~ &  & ~ &  & ~ &  & ~ &  & ~ &  & ~ &  & 34.74 & 3.88 & ~ &  & ~ &  \\ \hline 
42.8 & 1.8 &  & ~ &  & ~ &  & ~ &  & ~ &  & ~ &  & ~ &  & 27.61 & 3.01 & ~ &  & ~ &  \\ \hline 
\end{tabular}

\end{center}
\end{table*}

\begin{table*}[t]
\tiny
\caption{Experimental cross sections of $^{nat}$Pd(p,xn)$^{99m,99g,100,101m,101g,102m,102g,105}$Rh and $^{103}$Ru reactions}
\begin{center}
\begin{tabular}{|p{0.1in}|p{0.1in}|p{0.1in}|p{0.2in}|p{0.1in}|p{0.2in}|p{0.1in}|p{0.2in}|p{0.1in}|p{0.2in}|p{0.1in}|p{0.2in}|p{0.1in}|p{0.2in}|p{0.1in}|p{0.2in}|p{0.1in}|p{0.2in}|p{0.1in}|p{0.2in}|p{0.1in}|} \hline 
\multicolumn{2}{|c|}{\textbf{E$\pm\Delta$}E (MeV)} & \textbf{Exp.} & \multicolumn{18}{|c|}{\textbf{$\sigma \pm \Delta\sigma$} (mb)} \\ \hline 
\multicolumn{2}{|c|}{ } &  & \multicolumn{2}{|c|}{\textbf{$^{105}$Rh}} & \multicolumn{2}{|c|}{\textbf{$^{102m}$Rh}} & \multicolumn{2}{|c|}{\textbf{$^{102g}$Rh}} & \multicolumn{2}{|c|}{\textbf{$^{101m}$Rh}} & \multicolumn{2}{|c|}{\textbf{$^{101g}$Rh}} & \multicolumn{2}{|c|}{\textbf{$^{100}$Rh}} & \multicolumn{2}{|c|}{\textbf{$^{99m}$Rh}} & \multicolumn{2}{|c|}{\textbf{$^{99g}$Rh}} & \multicolumn{2}{|c|}{\textbf{$^{103}$Ru}} \\ \hline 
15.8 & 0.5 & V1 & 4.53 & 0.51 & ~ &  &  &  & 3.14 & 0.35 & 4.25 & 1.02 & ~ &  & ~ &  & ~ &  & ~ &  \\ \hline 
14.9 & 0.5 &  & 4.40 & 0.50 & ~ &  &  &  & 1.87 & 0.21 & ~ &  & ~ &  & ~ &  & ~ &  & ~ &  \\ \hline 
14.0 & 0.5 &  & 3.92 & 0.45 & ~ &  &  &  & 1.00 & 0.12 & ~ &  & ~ &  & ~ &  & ~ &  & ~ &  \\ \hline 
13.0 & 0.6 &  & 3.42 & 0.39 & ~ &  &  &  & 0.68 & 0.08 & ~ &  & ~ &  & ~ &  & ~ &  & ~ &  \\ \hline 
11.9 & 0.6 &  & 2.65 & 0.31 & ~ &  &  &  & 0.53 & 0.07 & ~ &  & ~ &  & ~ &  & ~ &  & ~ &  \\ \hline 
10.8 & 0.6 &  & 2.04 & 0.24 & ~ &  &  &  & 0.28 & 0.04 & ~ &  & ~ &  & ~ &  & ~ &  & ~ &  \\ \hline 
9.5 & 0.7 &  & 1.74 & 0.20 & ~ &  &  &  & 0.25 & 0.03 & ~ &  & ~ &  & ~ &  & ~ &  & ~ &  \\ \hline 
8.2 & 0.8 &  & 1.05 & 0.13 & ~ &  &  &  & 0.11 & 0.02 & ~ &  & ~ &  & ~ &  & ~ &  & ~ &  \\ \hline 
6.7 & 0.9 & ~ & 0.54 & 0.06 & ~ & ~ & ~ & ~ & 0.03 & 0.01 & ~ & ~ & ~ & ~ & ~ & ~ & ~ & ~ & ~ & ~ \\ \hline 
28.2 & 0.5 & V2 & 3.80 & 0.44 & ~ &  & ~ &  & 18.30 & 2.06 & 6.77 & 0.98 & 8.81 & 0.99 & ~ &  & ~ &  & ~ &  \\ \hline 
26.6 & 0.5 &  & 4.62 & 0.54 & ~ &  & ~ &  & 19.65 & 2.21 & 5.12 & 0.85 & 6.64 & 0.75 & ~ &  & ~ &  & ~ &  \\ \hline 
25.0 & 0.5 &  & 5.41 & 0.62 & ~ &  & ~ &  & 18.60 & 2.09 & ~ &  & 4.91 & 0.55 & ~ &  & ~ &  & ~ &  \\ \hline 
23.7 & 0.6 &  & 6.39 & 0.73 & ~ &  & ~ &  & 18.53 & 2.08 & 3.40 & 0.62 & 4.04 & 0.46 & ~ &  & ~ &  & ~ &  \\ \hline 
22.4 & 0.6 &  & 6.00 & 0.68 & ~ &  & ~ &  & 16.56 & 1.86 & ~ &  & 1.19 & 0.13 & ~ &  & ~ &  & ~ &  \\ \hline 
21.1 & 0.7 &  & 7.09 & 0.81 & ~ &  & ~ &  & 17.10 & 1.92 & ~ &  & 2.01 & 0.23 & ~ &  & ~ &  & ~ &  \\ \hline 
19.7 & 0.8 &  & 7.30 & 0.83 & ~ &  & ~ &  & 13.81 & 1.55 & ~ &  & 0.89 & 0.11 & ~ &  & ~ &  & ~ &  \\ \hline 
18.2 & 0.9 &  & 6.85 & 0.78 & ~ &  & ~ &  & 10.51 & 1.18 & ~ &  & 0.19 & 0.03 & ~ &  & ~ &  & ~ &  \\ \hline 
16.5 & 1.0 &  & 6.55 & 0.74 & ~ &  & ~ &  & 5.89 & 0.66 & ~ &  & ~ &  & ~ &  & ~ &  & ~ &  \\ \hline 
15.9 & 0.8 & ~ & 5.91 & 0.67 & ~ & ~ & ~ & ~ & 4.25 & 0.48 & ~ & ~ & ~ & ~ & ~ & ~ & ~ & ~ & ~ & ~ \\ \hline 
37.0 & 0.5 & V3 & 5.53 & 0.64 & ~ &  & 11.31 & 1.43 & 22.65 & 2.55 & 6.95 & 1.42 & 20.10 & 2.28 & 5.95 & 0.69 & 2.42 & 0.30 & ~ &  \\ \hline 
36.2 & 0.5 &  & 4.66 & 0.53 & ~ &  & 13.26 & 1.61 & 17.05 & 1.92 & 6.84 & 1.37 & 15.42 & 1.74 & 4.43 & 0.51 & 1.59 & 0.21 & ~ &  \\ \hline 
35.4 & 0.5 &  & 5.92 & 0.68 & ~ &  &  &  & 21.53 & 2.42 & 8.44 & 1.65 & 20.82 & 2.36 & 5.17 & 0.61 & 2.15 & 0.27 & ~ &  \\ \hline 
34.5 & 0.5 &  & 4.98 & 0.57 & ~ &  &  &  & 18.38 & 2.06 & 6.81 & 1.47 & 17.68 & 2.00 & 3.93 & 0.46 & 1.92 & 0.25 & ~ &  \\ \hline 
33.6 & 0.6 &  & 4.65 & 0.54 & ~ &  &  &  & 18.86 & 2.12 & 5.93 & 1.42 & 17.17 & 1.95 & 3.89 & 0.46 & 1.26 & 0.19 & ~ &  \\ \hline 
32.7 & 0.6 &  & 4.00 & 0.47 & ~ &  &  &  & 16.71 & 1.88 & 7.88 & 1.53 & 14.84 & 1.68 & 2.73 & 0.32 & 1.04 & 0.17 & ~ &  \\ \hline 
31.9 & 0.7 &  & 4.12 & 0.47 & ~ &  &  &  & 18.51 & 2.08 & ~ &  & 15.04 & 1.70 & 2.14 & 0.27 & 0.78 & 0.15 & ~ &  \\ \hline 
30.9 & 0.7 &  & 4.06 & 0.47 & ~ &  &  &  & 19.50 & 2.19 & ~ &  & 14.07 & 1.59 & 1.77 & 0.23 & 0.35 & 0.11 & ~ &  \\ \hline 
30.0 & 0.8 &  & 3.96 & 0.47 & ~ &  &  &  & 19.09 & 2.15 & ~ &  & 11.87 & 1.35 & 0.90 & 0.14 & ~ &  & ~ &  \\ \hline 
29.0 & 0.7 &  & 3.57 & 0.42 & ~ &  &  &  & 18.97 & 2.13 & ~ &  & 9.62 & 1.10 & 0.60 & 0.10 & ~ &  & ~ &  \\ \hline 
28.0 & 0.5 &  & 3.41 & 0.40 & ~ &  &  &  & 18.71 & 2.10 & ~ &  & 7.62 & 0.88 & 0.35 & 0.10 & ~ &  & ~ &  \\ \hline 
27.0 & 0.6 &  & 3.54 & 0.42 & ~ &  &  &  & 16.55 & 1.86 & ~ &  & 5.64 & 0.66 & ~ &  & ~ &  & ~ &  \\ \hline 
25.9 & 0.7 &  & 4.28 & 0.50 & ~ &  &  &  & 19.86 & 2.23 & ~ &  & 5.80 & 0.67 & ~ &  & ~ &  & ~ &  \\ \hline 
24.8 & 0.8 &  & 4.31 & 0.49 & ~ &  &  &  & 16.53 & 1.86 & ~ &  & 4.21 & 0.48 & ~ &  & ~ &  & ~ &  \\ \hline 
23.7 & 1.0 &  & 4.94 & 0.57 & ~ &  &  &  & 16.88 & 1.90 & ~ &  & 3.68 & 0.43 & ~ &  & ~ &  & ~ &  \\ \hline 
22.5 & 1.1 &  & 5.60 & 0.64 & ~ &  &  &  & 16.72 & 1.88 & ~ &  & 3.38 & 0.40 & ~ &  & ~ &  & ~ &  \\ \hline 
21.4 & 1.3 &  & 5.78 & 0.65 & ~ &  &  &  & 15.63 & 1.76 & ~ &  & 2.00 & 2.01 & ~ &  & ~ &  & ~ &  \\ \hline 
20.1 & 1.4 & ~ & 5.68 & 0.64 & ~ & ~ &  &  & 13.40 & 1.51 & ~ & ~ & 0.92 & 0.12 & ~ & ~ & ~ & ~ & ~ & ~ \\ \hline 
79.2 & 0.5 & S2004 & 12.58 & 1.80 & 31.20 & 10.35 & 23.46 & 3.26 & 182.86 & 20.53 & 15.39 & 6.23 & 55.22 & 6.20 & 37.20 & 4.19 & 7.28 & 0.88 & 1.37 & 0.25 \\ \hline 
74.9 & 0.5 &  & 11.81 & 1.37 & 26.58 & 8.92 & 18.78 & 2.53 & 156.44 & 17.56 & 21.38 & 5.74 & 41.37 & 4.65 & 25.87 & 2.91 & 5.71 & 0.70 & 1.19 & 0.22 \\ \hline 
70.4 & 0.6 &  & 10.90 & 1.30 & 22.82 & 10.73 & 19.82 & 2.88 & 158.23 & 17.76 & 11.40 & 3.72 & 43.88 & 4.93 & 27.04 & 3.05 & 5.35 & 0.65 & 0.86 & 0.17 \\ \hline 
65.7 & 0.8 &  & 8.82 & 1.05 & 23.78 & 13.09 & 26.39 & 3.43 & 184.09 & 20.67 & ~ &  & 34.51 & 3.88 & 24.72 & 2.79 & 6.33 & 0.77 & 1.13 & 0.22 \\ \hline 
60.7 & 1.0 &  & 10.20 & 1.17 & 24.51 & 7.54 & 31.09 & 3.98 & 168.28 & 18.89 & ~ &  & 27.99 & 3.14 & 23.83 & 2.68 & 6.38 & 0.76 & 0.87 & 0.19 \\ \hline 
56.5 & 1.2 &  & 7.27 & 0.88 & 19.53 & 7.09 & 27.96 & 3.57 & 123.82 & 13.90 & 6.97 & 3.70 & 24.05 & 2.70 & 20.24 & 2.28 & 5.13 & 0.61 & 0.65 & 0.17 \\ \hline 
52.0 & 1.4 &  & 6.23 & 0.78 & 14.98 & 9.42 & 27.73 & 3.41 & 95.23 & 10.69 & 8.63 & 4.54 & 23.62 & 2.65 & 19.43 & 2.19 & 5.07 & 0.59 & 0.61 & 0.13 \\ \hline 
47.2 & 1.1 &  & 7.65 & 0.88 & 7.44 & 1.79 & 23.51 & 2.99 & 67.12 & 7.54 & ~ &  & 22.32 & 2.51 & 15.66 & 1.77 & 4.20 & 0.50 & 0.51 & 0.14 \\ \hline 
42.0 & 1.5 &  & 6.71 & 0.78 & 5.06 & 1.46 &  &  & 29.10 & 3.27 & 8.16 & 4.91 & 19.62 & 2.21 & 6.94 & 0.79 & 2.62 & 0.37 & 0.34 & 0.16 \\ \hline 
36.2 & 1.9 & ~ & 3.94 & 0.44 & ~ & ~ &  &  & 21.61 & 2.43 & ~ & ~ & 13.62 & 1.53 & 2.27 & 0.28 & 0.63 & 0.22 & ~ & ~ \\ \hline 
67.7 & 0.5 & S2003 & ~ &  & ~ &  & ~ &  & ~ &  & ~ &  & 30.10 & 9.79 & ~ &  & 4.67 & 1.41 & ~ &  \\ \hline 
63.2 & 0.5 &  & ~ &  & ~ &  & ~ &  & ~ &  & ~ &  & 23.34 & 3.42 & ~ &  & ~ &  & ~ &  \\ \hline 
58.4 & 0.6 &  & ~ &  & ~ &  & ~ &  & ~ &  & ~ &  & 22.15 & 3.18 & ~ &  & ~ &  & ~ &  \\ \hline 
53.2 & 0.8 &  & ~ &  & ~ &  & ~ &  & ~ &  & ~ &  & 23.34 & 3.62 & ~ &  & ~ &  & ~ &  \\ \hline 
47.6 & 1.0 &  & ~ &  & ~ &  & ~ &  & ~ &  & ~ &  & 20.46 & 3.13 & ~ &  & ~ &  & ~ &  \\ \hline 
41.3 & 1.2 &  & ~ &  & ~ &  & ~ &  & ~ &  & ~ &  & 17.57 & 2.97 & ~ &  & ~ &  & ~ &  \\ \hline 
34.1 & 1.6 & ~ & ~ & ~ & ~ & ~ & ~ & ~ & ~ & ~ & ~ & ~ & 10.19 & 2.55 & ~ & ~ & ~ & ~ & ~ & ~ \\ \hline 
69.5 & 0.5 & S2002-1 & 16.34 & 2.93 & 20.48 & 9.18 & 26.83 & 4.92 & ~ &  & ~ &  & 47.96 & 5.24 & 22.79 & 3.74 & 6.99 & 0.78 & ~ &  \\ \hline 
65.5 & 0.5 &  & 15.20 & 2.79 & 12.25 & 7.61 & 26.83 & 4.89 & ~ &  & ~ &  & 39.57 & 4.33 & 20.49 & 3.36 & 6.58 & 0.93 & 1.16 & 0.35 \\ \hline 
61.3 & 0.6 &  & 9.92 & 2.08 & 15.03 & 8.84 & 23.53 & 4.48 & ~ &  & ~ &  & 32.47 & 3.56 & 18.41 & 3.19 & 5.15 & 0.75 & 0.67 & 0.39 \\ \hline 
56.9 & 0.7 &  & 12.03 & 2.52 & 17.96 & 9.30 & 23.10 & 4.37 & ~ &  & ~ &  & 27.92 & 3.08 & 20.37 & 3.62 & 4.27 & 0.66 & ~ &  \\ \hline 
52.2 & 0.9 &  & 10.48 & 2.06 & 16.79 & 8.79 & 22.46 & 4.39 & ~ &  & ~ &  & 26.00 & 2.86 & 20.06 & 3.28 & 4.58 & 0.65 & ~ &  \\ \hline 
47.2 & 1.1 &  & 10.58 & 1.97 & 13.56 & 8.80 & 22.25 & 4.36 & ~ &  & ~ &  & 23.04 & 2.54 & 19.71 & 3.23 & 3.18 & 0.60 & ~ &  \\ \hline 
41.8 & 1.4 &  & 9.27 & 2.04 & 8.25 & 8.05 & 12.92 & 2.98 & ~ &  & ~ &  & 20.40 & 2.26 & 9.66 & 2.42 & 1.80 & 0.57 & ~ &  \\ \hline 
35.8 & 1.8 &  & 7.83 & 2.12 & ~ &  & 10.73 & 2.83 & ~ &  & ~ &  & 18.03 & 2.02 & 6.46 & 2.56 & ~ &  & ~ &  \\ \hline 
28.9 & 2.2 & ~ & 3.83 & 1.64 & ~ & ~ &  &  & ~ & ~ & ~ & ~ & 8.35 & 0.98 & ~ & ~ & ~ & ~ & ~ & ~ \\ \hline 
67.4 & 0.5 & S2002-2 & 12.18 & 3.47 & 15.10 & 2.53 & 27.65 & 2.03 & ~ &  & ~ &  & 33.91 & 3.69 & 23.38 & 2.56 & 4.31 & 0.76 & 1.30 & 0.39 \\ \hline 
64.6 & 0.5 &  & 13.75 & 3.93 & 14.65 & 2.64 & 24.61 & 1.81 & ~ &  & ~ &  & 29.13 & 3.19 & 21.36 & 2.35 & 4.94 & 0.84 & 0.75 & 0.44 \\ \hline 
61.8 & 0.6 &  & 10.34 & 3.20 & 16.61 & 2.91 & 27.13 & 1.98 & ~ &  & ~ &  & 26.30 & 2.88 & 20.16 & 2.23 & 5.28 & 0.92 & ~ &  \\ \hline 
58.8 & 0.7 &  & 11.34 & 3.36 & 14.10 & 2.97 & 23.52 & 1.75 & ~ &  & ~ &  & 24.63 & 2.70 & 19.85 & 2.20 & 4.46 & 0.68 & ~ &  \\ \hline 
55.7 & 0.8 &  & 8.11 & 3.31 & 15.70 & 2.83 & 22.81 & 1.71 & ~ &  & ~ &  & 24.33 & 2.67 & 20.26 & 2.24 & 5.50 & 0.82 & ~ &  \\ \hline 
52.5 & 0.9 &  & 8.48 & 3.12 & 13.98 & 2.71 & 22.15 & 1.65 & ~ &  & ~ &  & 23.91 & 2.63 & 18.47 & 2.05 & 4.71 & 0.71 & ~ &  \\ \hline 
49.1 & 1.1 &  & 8.00 & 3.01 & 12.02 & 2.38 & 19.81 & 1.58 & ~ &  & ~ &  & 22.60 & 2.48 & 17.06 & 1.90 & 3.84 & 0.66 & ~ &  \\ \hline 
45.6 & 1.3 &  & 4.12 & 2.84 & 6.14 & 2.29 & 16.92 & 1.34 & ~ &  & ~ &  & 21.16 & 2.33 & 14.99 & 1.68 & 3.15 & 0.63 & ~ &  \\ \hline 
44.2 & 1.6 &  & 5.45 & 2.75 & 8.86 & 2.84 & 13.55 & 1.19 & ~ &  & ~ &  & 20.48 & 2.26 & 13.78 & 1.56 & 2.89 & 0.53 & ~ &  \\ \hline 
42.8 & 1.8 &  & 4.78 & 1.45 & 5.75 & 2.20 & 11.61 & 1.08 & ~ &  & ~ &  & 19.50 & 2.12 & 11.61 & 1.27 & 2.67 & 0.31 & ~ &  \\ \hline 
\end{tabular}
\end{center}
\end{table*}

\subsection{Integral yields}
\label{3.2}

\noindent The integral yields for $^{103,104m,104g,105g,106m,110m}$Ag, $^{100,101}$Pd, $^{99m,100mg,100,101m,101g,102m,102g,105}$Rh and $^{97,103}$Ru reactions were calculated from spline fits to our experimental excitation functions. The integral yields represent so called physical yields i.e. activity instantaneous production rates \citep{Bonardi, Otuka}. The integral yields are shown In Figs. 19 and 20 with experimental thick target yield data found in the literature for the comparison. Only one previous experimental data-set was found in the literature \citep{Hermanne2004b}, which shows good agreement with our results (see Fig. 19). On Figs. 19-20 the yield of different isotopes covers a range of 5-6 orders of magnitude, so it is easy to determine, which radio-isotope is economic for production. Out of the production parameters also the radionuclidic purity of the product must be investigated.

\begin{figure}
\includegraphics[width=0.5\textwidth]{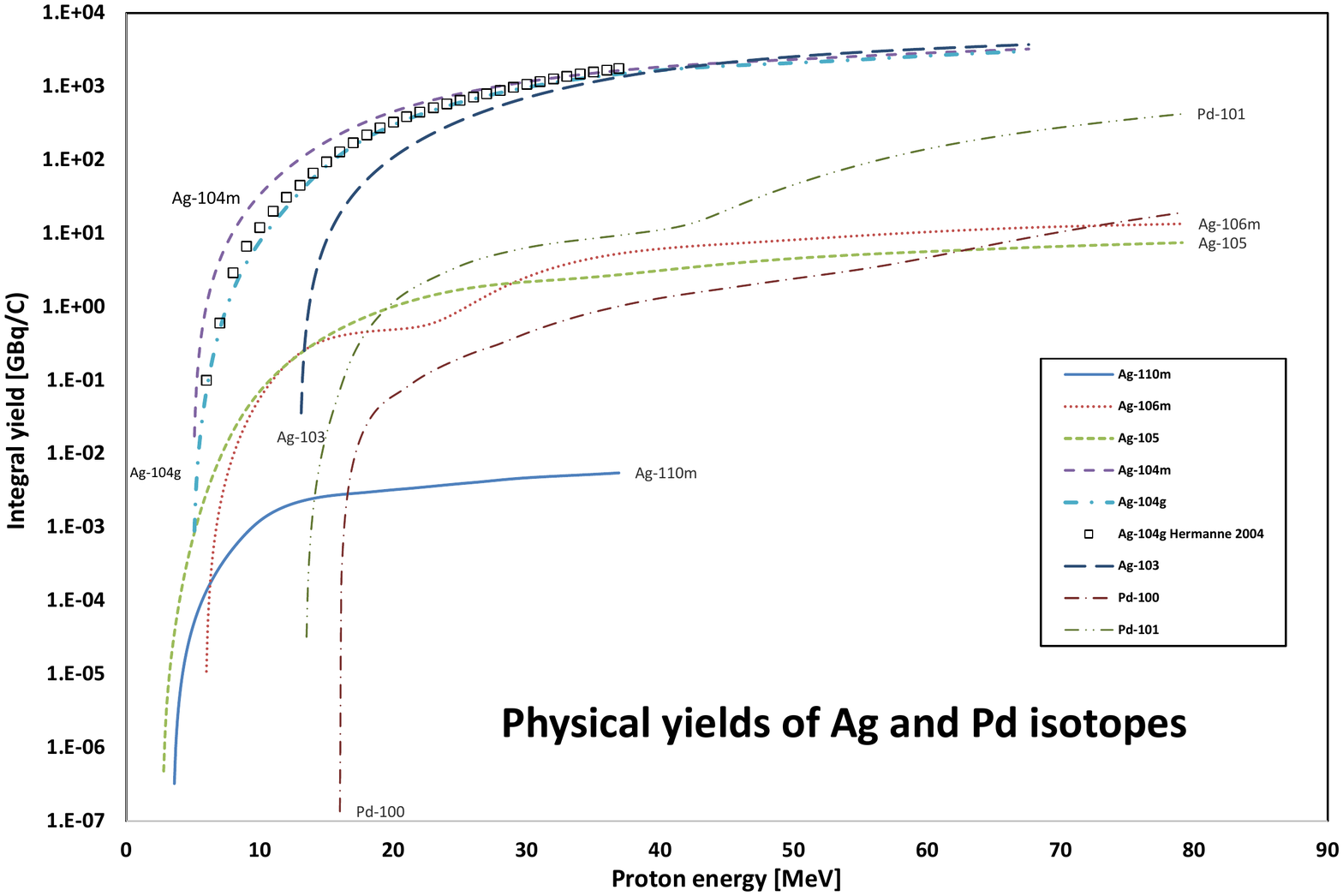}
\caption{Integral yields for the~$^{103,104m,104g,105g,106m,110m}$Ag and $^{100,101}$Pd producing reactions}
\label{fig:19}       
\end{figure}

\begin{figure}
\includegraphics[width=0.5\textwidth]{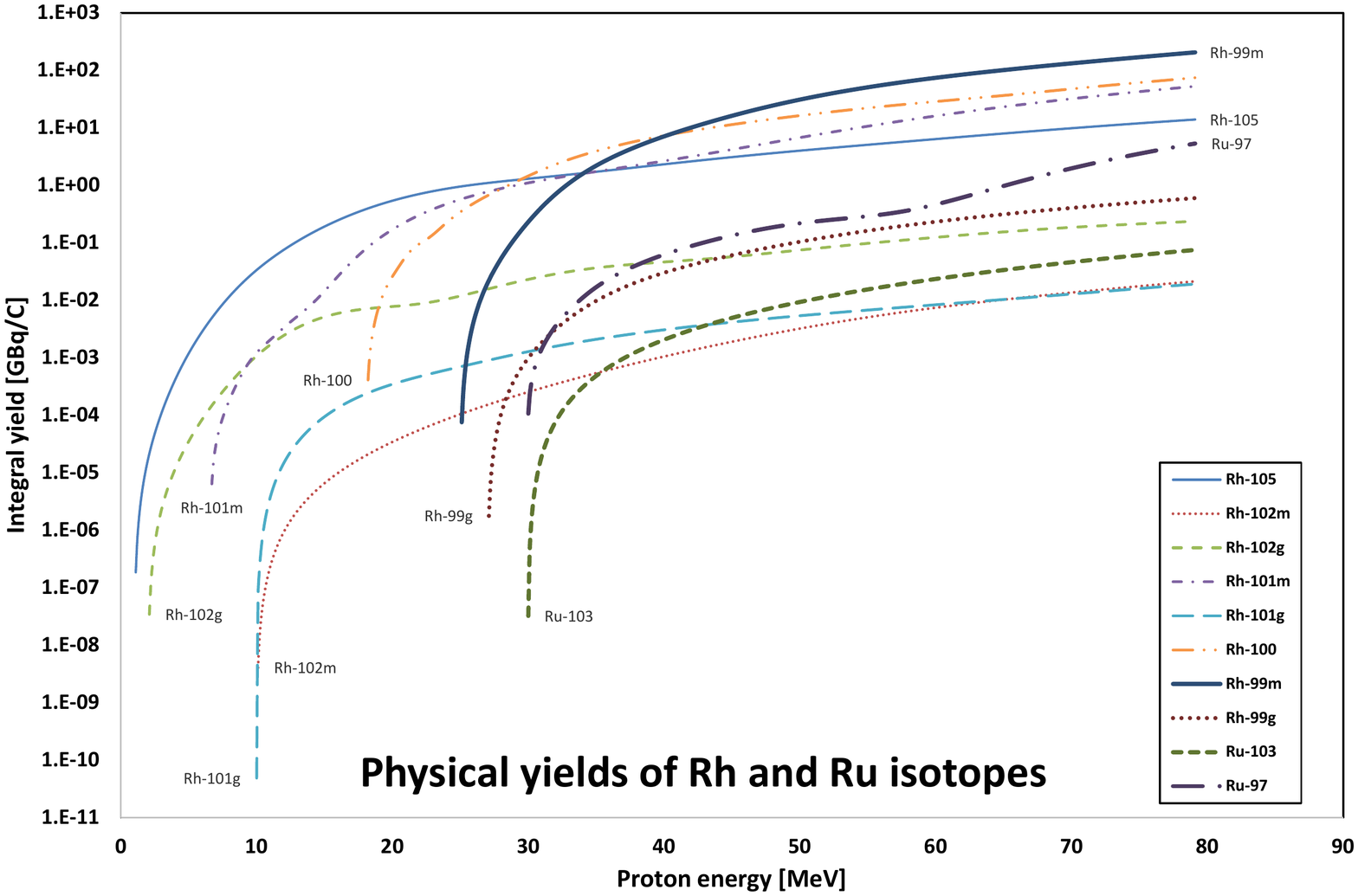}
\caption{Integral yields for the~$^{99m,100,101m,101g,102m,102g,105}$Rh and $^{97,103}$Ru  producing reactions}
\label{fig:20}       
\end{figure}

\section{Overview of production routes of assessed activation products}
\label{4}

\noindent Many of the investigated radio-products of proton induced nuclear reactions on palladium have applications in different fields of biological studies, nuclear medicine and labeling for industrial processes. Detailed comparison of different production routes would require an extended review. Here we summarize only a few application fields of these isotopes. 

\noindent \textbf{\textit{$^{103}$Pd (}}through\textbf{\textit{ $^{103}$Ag) }}Because of its suitable half-life and decay characteristics the $^{103}$Pd (16.991 d) is clinically used in permanent brachytherapy \citep{Bakhshabadi}.  For a review of production routes of ${}^{103}$Pd we refer to our previous work \citep{TF2009}. According to this review: ``At medium and low energy accelerators (up to 100 MeV) ${}^{103}$Pd can be directly produced through the $^{101}$Ru($^{3}$He,n), ${}^{102}$Ru($^{3}$He,2n), $^{100}$Ru($\alphaup$,n), ${}^{101}$Ru($\alphaup$,2n), ${}^{102}$Ru($\alphaup$,3n), $^{103}$Rh(p,n), ${}^{103}$Rh(d,2n) and $^{nat}$Ag(p,x) reactions. Also indirect production is possible by the decay of ${}^{103}$Ag (T${}_{1/2}$= 65.7 min). Parent $^{103}$Ag can be produced via ${}^{103}$Rh($\alphaup$,4n), $^{nat}$Pd(p,x), ${}^{104}$Pd(p,2n), ${}^{nat}$Pd(d,x),${}^{ nat}$Pd($\alphaup$,x), $^{nat}$Cd(p,x), $^{106}$Cd(p,x), $^{nat}$Cd(d,x) and $^{106}$Cd(d,x) processes.

\noindent The preferential routes in different energy domains are the following: at low energy accelerators (E$_{p,max}$ = 20 MeV) direct production routes via the $^{103}$Rh(p,n), $^{103}$Rh(d,2n) and indirect route via the $^{104}$Pd(p,2n) reaction can be considered. Among these the widely used $^{103}$Rh(p,n) process has the highest yields.

  At medium energy cyclotrons (E$_{p,max}$ = 30-50 MeV) the $^{104}$Pd(p,2n) and $^{nat}$Pd(p,x) routes can be considered in addition to the previously mentioned possibilities. Out of them the $^{104}$Pd(p,2n) and $^{103}$Rh(d,2n) processes have the highest yields. If even higher energies are available (E$_{p,max }$$\sim$100 MeV), the indirect production routes become more productive. The $^{106}$Cd(p,x) and $^{104}$Pd(p,2n) processes have the highest yields but the ${}^{nat}$Ag(p,x), $^{nat}$Pd(p,x) and ${}^{nat}$Cd(p,x) reactions are the most promising because of the cheap metal targets of natural isotopic composition. The production yields of these high energy routes are one order of magnitude higher than the presently used $^{103}$Rh(p,n) reaction, but the using of higher energy introduces further impurities into the final product. Replacement of 10 small cyclotrons (18 MeV protons used for $^{103}$Pd production) with a high energy one having comparable beam intensity could hence be an option that should be investigated as it would allow also parallel production of other radioisotopes from different target materials needing proton energies around 30 MeV (e.g. $^{67,68}$Ga, $^{203}$Tl, $^{111}$In, $^{103}$Pd on Rh and $^{99}$Mo)'', see also \citep{Adam, TF2011}.

\noindent The \textbf{\textit{$^{104g}$Ag}} (T$_{1/2}$ = 69.2 min, ${\beta}^{+}$: 15 \%) is a PET imaging analogue of the therapeutic radionuclide $^{111}$Ag. Production routes were discussed in \citep{Hermanne2004b}. Nearly pure $^{104}$Ag (m+g) can be produced with 15 MeV protons on highly enriched $^{104}$Pd targets. A possible drawback for quantitative PET imaging is the changing contribution of the metastable state with relatively high $\betaup^{+}$ contribution and shorter half-life. Independent cross section determination and evaluation of contribution is possible but asks for specially planned experiment with several fast measurements of the decay in the first hours after EOB.

\noindent Contamination with $^{104}$Pd(d,n)$^{105}$Ag (T$_{1/2}$ = 41.3 d) reaction cannot be avoided when using deuteron induced reactions for $^{104}$Ag production. As the values of the cross sections for proton and deuteron induced reactions are similar but yields for the proton induced reaction are higher due to the longer range, the $^{104}$Pd(p,n)$^{104m,g}$Ag reaction on targets of 500 mg/cm$^{2}$ and with incident energy of E$_{p} <$ 15 MeV has to be preferred.

\noindent The\textbf{\textit{ $^{105,106}$Ag}} longer lived radioisotopes of silver are widely use to study Ag metabolism, labeling of nanoparticles, diffusion in solids, etc. The thin layer activation study of palladium the $^{105}$Ag product through $^{nat}$Pd(p,x)$^{105}$Ag is used \citep{DF2012}.

\noindent At low and medium energy accelerators the $^{105}$Ag (T$_{1/2}$ = 41.29 d) can be produced via $^{105}$Pd(p,n), $^{104}$Pd(d,n)  and  $^{103}$Rh($\alpha$,2n) and $^{103}$Rh($^{3}$He,n) reactions. Out of them the $^{105}$Pd(p,n) is the most productive. In case of alpha-particle beam the $^{106}$Ag is also always produced simultaneously via (?,n).

\noindent Similarly for production of $^{106}$Ag (T$_{1/2}$ = 8.28 d) the $^{106}$Pd(p,n), $^{106}$Pd(d,2n) and $^{103}$Rh(?,n) reactions are available. In this mass region the yield of the (d,2n) is often double of that of the (p,n) reaction on the same target, but higher energy cyclotrons are needed. The availability of high intensity alpha particle beam is low at commercial cyclotrons and the yield is low, due to the high stopping for alpha particles.

\noindent The \textbf{\textit{$^{110m}$Ag}} was used for silver uptake studies in different biological organisms \citep{Zalewska}, for characterization of [$^{110m}$Ag]-nanoparticles in different organisms, industrial tracing of silver in long term processes, etc.  It can be produced via $^{110}$Pd(p,n) and $^{110}$Pd(d,2n) reaction. In case of deuteron route the $^{111}$Ag is produced simultaneously via (d,n) reaction, but it practically  does not disturb the above mentioned tracing applications. The (d,2n) requires higher energy cyclotron, the maximum of (d,2n) is shifted above 6 MeV to higher energies. 

\noindent The\textbf{\textit{ $^{105}$Rh }}(T$_{1/2}$ = 35.36 h)\textbf{\textit{, $^{101m}$Rh }}(T$_{1/2}$ = 4.34 d, through $^{101}$Pd),\textbf{\textit{$^{101}$Rh }}(3.3 a),\textbf{\textit{ $^{97}$Ru}}(T$_{1/2}$ = 2.83 d) have been considered as potential candidates for targeted radio-therapeutic use \citep{Pakravan}. For production of these radioisotopes the proton induced reactions on palladium play smaller roles. Their production is more advantageous by light charged particle induced reactions on isotopes of ruthenium. 

\noindent Shortly summarizing: out of the above mentioned radio-products - taking into account different reaction routes - the proton induced reactions on stable isotopes of palladium are more competitive for production of silver isotopes (availability of beams, high cross sections of (p,xn) reactions, etc.).

\section{Summary}
\label{5}

\noindent Excitation functions for direct and cumulative cross-sections for the formation of $^{103,104m,104g,105g,106m,110m}$Ag, $^{100,101}$Pd, $^{99m,100,101m,101g,102m,102g105}$Rh and $^{103,97}$Ru radioisotopes  on palladium have been measured up to 80 MeV proton energy.  Experimental activation cross-section data are presented for the first time in the 70-80 MeV region for all measured isotopes, for $^{104m}$Ag, $^{104g}$Au, $^{101}$Pd, $^{100}$Rh and $^{105}$Rh in the 40-80 MeV energy range, and for $^{102m}$Rh, $^{102g}$Rh, $^{101g}$Rh, $^{99m}$Rh, $^{99g}$Rh, $^{103}$Ru and $^{97}$Ru in the whole energy range. The agreement with the earlier experimental data in the overlapping regions is acceptable good, except one or two cases. The comparison with the theoretical predictions in the TENDL libraries show acceptable agreement, but in several cases, especially by some Ru and Rh ($^{103}$Ru and $^{102g,101g}$Rh) radioisotopes, where a continuous feeding from the excited state is possible, systematic underestimation can be observed. Because of the bad statistic by several measurement series large errors and big fluctuations can be observed, especially at higher energy experiments and by low activities, that's why further measurements are encouraged to clarify these cases.

\section{Acknowledgements}
\label{}
This work was performed in the frame of the HAS-FWO Vlaanderen (Hungary-Belgium) and HAS-JSPS (Hungary-Japan) projects. The authors acknowledge the support of the research project and of the respective institutions.
%\FloatBarrier
 


%% References

%\clearpage
\bibliographystyle{elsarticle-harv}
\bibliography{Pdp}






% \end{thebibliography}

\end{document}